\documentclass[letterpaper,twocolumn,10pt]{article}
\usepackage{usenix2019}
 \pdfoutput=1

\usepackage{tikz}
\usepackage{amsmath}
\usepackage[subtle]{savetrees}
\usepackage[small,compact]{titlesec}
\usepackage{filecontents}
\usepackage{xcolor}
\usepackage{xspace}
\usepackage{amsmath}
\usepackage{amsthm}
\usepackage{amsfonts}
\usepackage{amssymb}
\usepackage{subfigure}
\usepackage{multirow}
\usepackage{makecell}
\usepackage{wrapfig}
\usepackage[linesnumbered,boxed,ruled,noend,commentsnumbered,norelsize]{algorithm2e}
\usepackage{xurl}
\usepackage{booktabs}
\usepackage{enumitem}
\usepackage{graphics}
\usepackage{graphicx}
\usepackage{grffile}
\usepackage{siunitx}
\usepackage{tikz}
\usepackage{xcolor}
\usepackage{stix}
\usepackage{tabulary}
\usepackage{pifont}
\usepackage[font={small,bf}]{caption}
\usepackage{lastpage}

\newcommand*\circled[1]{\tikz[baseline=(char.base)]{
            \node[shape=circle,line width=0.75pt, draw=black,inner sep=1pt] (char) {\textcolor{black}{\small\textsf{#1}}};}}
\newcommand*\sfboxed[1]{\tikz[baseline=(char.base)]{
            \node[shape=rectangle,line width=0.75pt, draw=black,inner sep=2pt, rounded corners=2pt] (char) {\textcolor{black}{\small\textsf{#1}}};}}
\AtBeginDocument{%
  \providecommand\BibTeX{{%
    \normalfont B\kern-0.5em{\scshape i\kern-0.25em b}\kern-0.8em\TeX}}}

\newcommand{\vls}[1]{{\color{blue} \textbf{VLS:} {#1}}}

\newcommand{\fd}[1]{{\color{red} \textbf{FD:} {#1}}}
\newcommand{\sadjadf}[1]{{\color{purple} \textbf{SF:} {#1}}}

\newcommand{\todo}[1]{{\color{red} {#1}}}

\newcommand{\eg}{{\em e.g., }}

\newcommand{\etc}{{\em etc.}}
\newcommand{\vs}{{\em vs. }}

\newcommand{\Sec}[1]{\S\ref{#1}}
\newcommand{\App}[1]{App.~\ref{app:#1}}

\newcommand{\TheSystem}{Gemino\xspace}
\newcommand{\fom}{FOMM\xspace}
\newcommand{\kp}{keypoint\xspace}
\newcommand{\kps}{keypoints\xspace}
\newcommand{\aiortc}{\emph{aiortc}\xspace}
\newcommand{\vce}{video conference\xspace}

\newcommand{\vcing}{video conferencing\xspace}
\newcommand{\Vcing}{Video conferencing\xspace}
\newcommand{\bpp}{bits-per-pixel\xspace}
\newcommand{\ms}{multi-scale\xspace}
\newcommand{\PF}{PF\xspace}
\newcommand{\perframe}{per-frame\xspace}
\newcommand{\REF}{reference\xspace}

\newcommand{\Fig}[1]{Fig.~\ref{fig:#1}}
\newcommand{\NewPara}[1]{\noindent{\bf #1}}

\newcommand{\eat}[1]{}
\newcommand{\Tab}[1]{Tab.~\ref{tab:#1}\xspace}

\begin{document}

\date{}

\title{\Large \bf \TheSystem: Practical and Robust Neural Compression for Video Conferencing}


\author{
\rm
 \begin{tabulary}{\textwidth}{CCCC}
   Vibhaalakshmi Sivaraman & Pantea Karimi & Vedantha Venkatapathy & Mehrdad Khani \\
   Sadjad Fouladi~\raisebox{0.3ex}{\small$\boxplus$} & Mohammad Alizadeh & Fr\'edo Durand & Vivienne Sze
 \end{tabulary}\\[1.5ex  ]
 \textit{Massachusetts Institute of Technology}, \raisebox{0.3ex}{\small$\boxplus$}~\textit{Microsoft Research}
 \vspace{3ex}
 }

\maketitle


\begin{abstract}

\Vcing systems suffer from poor user experience when network conditions deteriorate because current video codecs simply cannot operate at extremely low bitrates. Recently, several neural alternatives have been proposed that reconstruct talking head videos at very low bitrates using sparse representations of each frame such as facial landmark information. However, these approaches produce poor reconstructions in scenarios with major movement or occlusions over the course of a call, and do not scale to higher resolutions. 
We design \TheSystem, a new neural compression system for \vcing based on a novel \emph{high-frequency-conditional super-resolution} pipeline. \TheSystem upsamples a very low-resolution version of each target frame while enhancing high-frequency details (e.g., skin texture, hair, etc.) based on information extracted from a single high-resolution reference image. 
We use a \ms architecture that runs different components of the model at different resolutions, allowing it to scale to resolutions comparable to 720p, and we personalize the model  to learn specific details of each person, achieving much better fidelity at low bitrates. We implement \TheSystem atop \aiortc, an open-source Python implementation of WebRTC, and show that it operates on 1024$\times$1024 videos  
 in real-time on a Titan X GPU, and achieves 2.2--5$\times$ lower bitrate than traditional video codecs for the same perceptual quality.

\end{abstract}


\section{Introduction}
\label{sec:intro}


\Vcing applications have become a crucial part of modern life.
However, today's systems continue to suffer from poor user experience: in particular, poor video quality and unwelcome disruptions are all too common. 
Many of these problems are rooted in the inability of today's applications 
to operate in low-bandwidth scenarios. For instance, Zoom recommends a minimum bandwidth of \SI{1.2}{Mbps} for one-on-one meetings and 2-3 Mbps for group meetings~\cite{zoom-mbps}. In certain parts of the world, Internet broadband speeds remain insufficient for reliable video conferencing. For example, large swaths of the population in Africa and Asia had average Internet broadband speeds less than 10 Mbps in 2022~\cite{broadband_speeds}, with the five slowest countries having speeds under 1 Mbps. Mobile bandwidth is even more restricted: SpeedTest's Global Index~\cite{speed_test} suggest that global mobile bandwidth average is 50\% of broadband speeds. Even in regions of North America and Europe with high broadband speeds~\cite{broadband_speeds}, over 30\% of users surveyed about their \vcing experience claimed that ``video quality'' issues were their biggest pain point~\cite{vc_experience}. This is because the user experience is not just determined by the average bandwidth, but rather by tail events of low bandwidth (a few seconds every 5--10 minutes) that cause glitches and disrupt the video call.
When the network deteriorates, even briefly, existing \vcing solutions cope to an extent by lowering quality, but below a certain bandwidth (e.g., 100s of Kbps for HD video), they must either suspend the transmission altogether or risk packet loss and frame corruption. 

Recently, several neural approaches for face image synthesis 
have been proposed that deliver extreme compression by generating each video frame from a sparse representation (\eg \kps)~\cite{fom,maxine,bilayer,fbfom,fom_multiple_views}. 
These techniques have the potential to enable \vcing with one to two orders of magnitude reduction in bandwidth (as low as \SI{\sim 10}{Kbps}~\cite{maxine,fbfom}), but their lack of robustness and high computational complexity hampers their practicality. Specifically, synthesis approaches work by ``warping'' a reference image into different target poses and orientations captured by such  sparse \kps. These methods produce good reconstructions when the difference between the reference and the target image is small, but they fail (possibly catastrophically) in the presence of large movements or occlusions. In such cases, they produce poor reconstructions, for both low-frequency content (\eg missing the presence of a hand in a frame altogether) and high-frequency content (\eg details of clothing and facial hair). As a result, while synthesis approaches show promising average-case behavior, their performance at the tail is riddled with inconsistencies in practice. Furthermore, real-time reconstruction is only feasible at low resolution for such models~\cite{fom,fbfom}, even on high-end GPUs, while typical \vce applications
are designed for HD, Full HD, and even 4K videos~\cite{trueconf, starline}.
Na\"ively reusing these models on larger input frames can quickly become prohibitively expensive as the resolution is increased.

\begin{figure}
\centering
    \includegraphics[width=\linewidth]{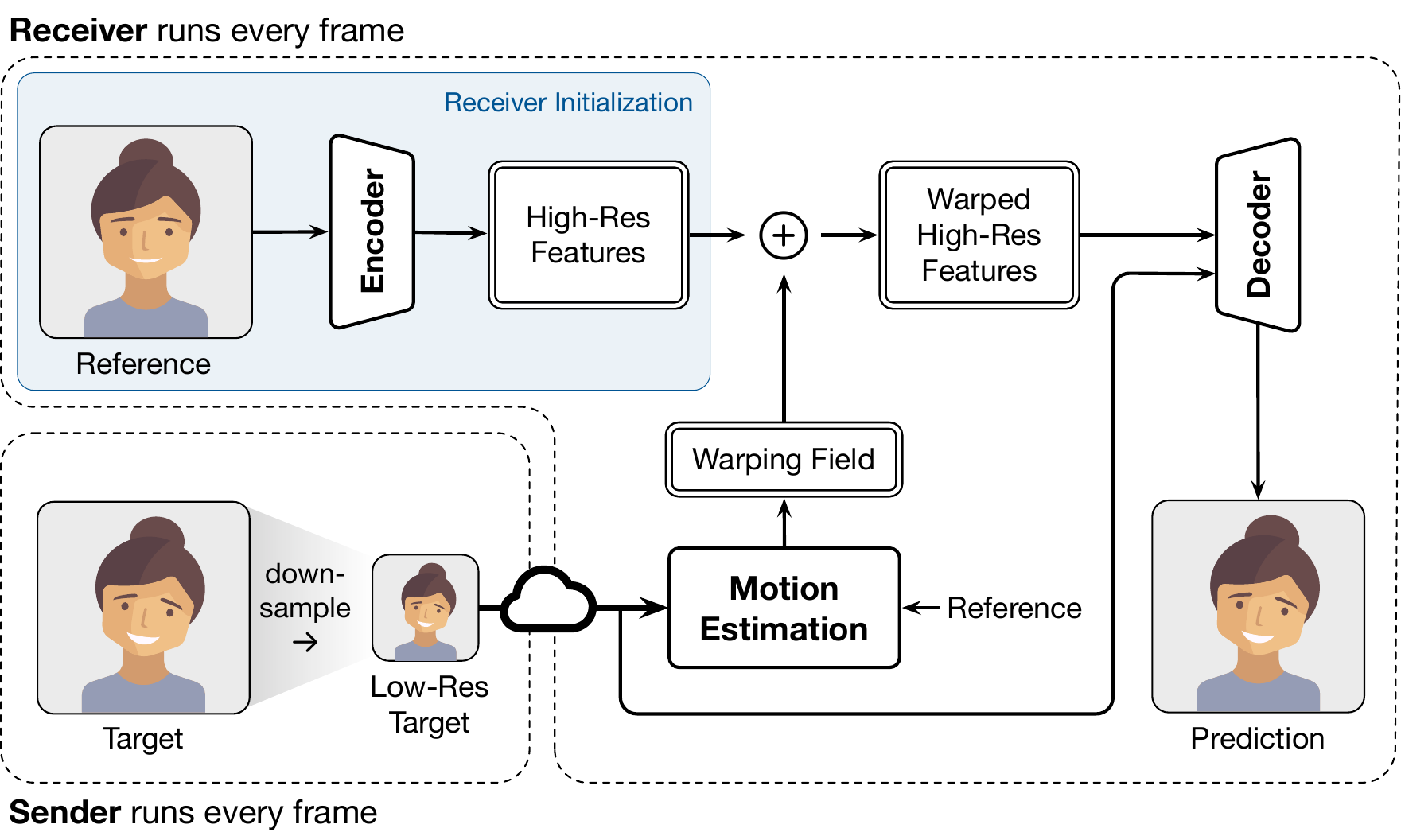} 
\caption{\small \TheSystem's design. The sender sends a downsampled version of the target frame across the network to the receiver that has a reference image of the speaker in the current \vce setting. The encoder network encodes and warps features from the reference image based on the motion estimated between the reference and target frames. 
The decoder network at the receiver upsamples the downsampled frame with help from the warped encoded reference features.}
\vspace{-4.5mm}
\label{fig:model structure}
\end{figure} 

We present \TheSystem, a neural compression system for low-bitrate \vcing, designed to overcome the above robustness and compute complexity challenges. Gemino targets extreme compression scenarios, such as delivering video in $\sim$100 Kbps or less. At such bitrates, the bandwidth required for video becomes comparable to a typical audio call~\cite{audio_recommentdation}, greatly expanding the range of networks that can support video conferencing. 

\TheSystem's design begins with the observation that current synthesis approaches, in an effort to squeeze the most compression, overly rely on \kps, a modality with limited semantic information about the target frame. This causes several inevitable failures (\Sec{sec:motivation}). For example, if a user's hand is absent in the reference frame but appears in the target, there is no way to reconstruct the hand by warping the reference image. We would need to send a new reference frame that includes the hand, but sending a high-resolution frame (even occasionally) incurs significant cost. Instead, \TheSystem directly transmits low-resolution video, which includes significantly more information about the target frame, and upsamples it to the desired resolution at the receiver. Using low-resolution video is viable because modern codecs~\cite{chen2018overview,mukherjee2015technical,bankoski2011technical,schwarz2007overview} compress them very efficiently. For example, sending 128$\times$128 resolution video in our implementation consumes $\sim$\SI{15}{Kbps}, only slightly more than would be needed to transmit \kps~\cite{maxine,fbfom}. We posit that the robustness benefits of providing the receiver with more information for reconstruction far outweigh the marginal bandwidth cost.

It is challenging to upsample a video significantly while reconstructing high-frequency details accurately. For example, at 1024$\times$1024 resolution, we can see high-frequency texture details of skin, hair, and clothing, that are not visible at 128$\times$128 resolution. To improve high-frequency reconstruction fidelity, \TheSystem uses a reference frame 
that provides such texture information in a different pose than the target frame. Like synthesis approaches, it warps features extracted from this reference frame based on the motion between the reference and target frames, but it combines it with information extracted from the low-resolution target image to generate the final reconstruction  (\Fig{model structure}). We call this new approach {\em high-frequency-conditional super-resolution}. 

\TheSystem uses several further optimizations to improve reconstruction fidelity and reduce computation cost. First, we \emph{personalize} the model by fine-tuning it on videos of a specific person so that it can learn the high-frequency detail associated with that person (e.g., hair, skin wrinkles, \etc). Personalizing the model makes it easier to transition texture information from the reference to the target frame. 
Second, we train the neural network using decompressed low-resolution frames obtained from a standard codec so that it learns to reconstruct accurate target frames despite codec-induced artifacts. We design \TheSystem to work with any starting resolution so that it can achieve different rate-distortion tradeoffs  based on the available network bandwidth. The resolution at a particular bitrate is chosen by profiling standard codecs and picking the highest resolution that can achieve that bitrate.


Lastly, as the target resolution increases, it is essential to reduce the number of operations required per-pixel for synthesis. Otherwise, the compute overheads of running neural networks at higher resolutions become prohibitively high. To achieve this compute reduction, we design a \emph{multi-scale} architecture wherein different parts of the model operate at different resolutions. For instance, the module that produces the warping field uses low-resolution versions of the reference and target images, while the encoder and decoder networks\footnote{``Encoder'' and ``decoder'' throughout this paper refer to the neural network pair typically used in GANs~\cite{goodfellow2014generative}. When we specify \emph{VPX} encoder or decoder, we refer to the video codec's encoder and decoder.} operate at full resolution but are equipped with additional downsampling blocks to reduce the operations per pixel. This design allows us to obtain good
reconstruction quality while keeping the reconstruction real time. The \ms architecture will be particularly salient as we transition to higher resolution \vcing applications in the future. We also further optimize our model using neural architecture search techniques such as NetAdapt~\cite{eccv2018-yang} to reduce the compute footprint.


We implement and evaluate \TheSystem within \aiortc~\cite{aiortc}, a Python implementation of WebRTC, and show the following: 
\begin{enumerate}[noitemsep,topsep=0pt,parsep=0pt,partopsep=0pt]
    \item \TheSystem achieves a perceptual quality (LPIPS)~\cite{lpips} of 0.21 at \SI{105}{Kbps}, a $5\times$ and $2.2\times$ reduction 
        from VP8 and VP9's default Chromium and WebRTC settings respectively. 
    \item In lower bitrate regimes where current \vcing applications \emph{cannot} operate, \TheSystem outperforms bicubic upsampling and SwinIR~\cite{liang2021swinir}  super-resolution
         from compressed downsampled VPX \footnote{VPX refers jointly to VP8 and VP9; we delineate them when necessary.} frames.
         \TheSystem requires \SI{50}{Kbps} to achieve
        an LPIPS of 0.23, a 2$\times$ and 3.9$\times$ reduction compared to bicubic and super-resolution. 
    \item Our model transitions smoothly across resolutions, and between generated and synthetic video, achieving different rate-distortion tradeoffs based on the target bitrate.
    \item Our model benefits considerably ($\sim$ 0.04 in LPIPS) 
    from personalization and including the VPX codec at train-time 
    ($\sim$0.05 in LPIPS).
    Our optimizations atop the \ms architecture enable real-time inference on 1024$\times$1024  frames on a Titan X GPU. 
    \end{enumerate}



\section{Related Work}
\label{sec:related}


\NewPara{Traditional Codecs}.
Most video applications rely on standard video compression modules (codecs) such as  H.264/H.265~\cite{h264, h265}, VP8/VP9~\cite{vp8, vp9}, and AV1~\cite{chen2018overview}.
These codecs separate video frames into keyframes (I-frames) that exploit spatial redundancies within a frame, and predicted frames (P-/B-frames) that exploit temporal---as well as spatial---redundancies across frames.
Over the years, these standards have been improved through ideas like variable block sizes~\cite{h265} and low-resolution encoding for lower bitrates~\cite{chen2018overview}. 
These codecs are particularly efficient in their \emph{slow} modes when they have generous time and compute budget to compress a video at high quality.
However, these codecs still require a few hundred Kbps for real-time applications such as \vcing, even at moderate resolutions like 720p.
In low-bandwidth scenarios, these codecs cannot do much other than transmit at the worst quality, and suffer packet loss and frame corruption~\cite{salsify}. To circumvent this, some applications~\cite{youtube} switch to lower resolutions when the network degrades. 
However, as new \vcing solutions such as Google's Starline~\cite{starline} with a large bandwidth footprint are introduced, these concerns with current codecs become more acute. 

\NewPara{Super-resolution.} 
Linear single-image super-resolution (SR) methods~\cite{bicubic, linear_interpolation} provide robust quality enhancements in various contexts. Neural SR methods have further enhanced the upsampling quality by learning better interpolation or in-painting methods~\cite{dong2014learning,esrgan,edsr,liang2021swinir}. 
Video SR methods~\cite{vespcn,vrt} build on image SR but further improve the reconstruction by exploiting redundant information in adjacent low-res video frames. Certain approaches like FAST~\cite{zhang2017fast} and Nemo~\cite{nemo} further optimize SR for video generation by performing SR only on ``anchor frames'' and generating the rest by upsampling motion vectors and residuals. For \vcing, domain-specific SR has also shown promising outcomes utilizing facial characteristics and training losses in their models~\cite{fsrnet,ma2020deep}. 
However, to the best of our knowledge, none of these prior methods study upsampling conditioned on a high-resolution image from the same context. 
Unlike pure SR methods, \TheSystem provides access to high-resolution reference frames and learns models that jointly in-paint and propagate high-frequency details from the reference frame. In recent work, SRVC~\cite{srvc} uses content-specific super-resolution to upsample a low-resolution video stream. Our approach is similar to SRVC in that it designs a model adapted to a specific person. However, to enable real-time encoding, \TheSystem only customizes the model once per person rather than continuously adapting it throughout the video.


\NewPara{Neural Codecs.}
The inability of traditional codecs to operate at extremely low bitrates for high-resolution videos has led researchers to consider neural approaches that reconstruct video from very compact representations.
Neural codecs have been designed for video-streaming~\cite{srvc,nas,swift}, live-video~\cite{livenas}, and video \vcing~\cite{maxine,fbfom}.
Swift~\cite{swift} learns to compress and decompress based on the residuals
in a layered-encoding stack. Both NAS~\cite{nas} and LiveNAS~\cite{livenas} enhance video quality using one or more DNN models at either the client for video streaming, or the ingest server
for live video. The models have knobs to control the compute overheads by using a smaller Deep Neural Network (DNN)~\cite{nas}, or by adjusting the number of epochs over which they are fine-tuned online~\cite{livenas}.
All of these approaches have shown improvements in the \bpp consumption across a wide range of videos. 

However, \vcing differs from other video applications in a few ways.
First, the video is unavailable ahead of time to optimize
for the best compression-quality tradeoff. Moreover, the interactivity of the application
demands that the video be both compressed and decompressed with low-latency.
Second, the videos belong to a specific distribution consisting primarily of facial data. 
This allows for a more targeted model for generating videos of faces.
A number of such models have been proposed~\cite{bilayer, maxine, fbfom, fom, dejavu, fom_multiple_views, implicit_warping} over the years. These models use \kps or facial landmarks as a compact intermediary representation of a specific pose, to compute the movement between two poses before generating the reconstruction. The models may use 3D \kps~\cite{maxine}, off-the-shelf \kp detectors~\cite{bilayer},
or multiple reference frames~\cite{fom_multiple_views} to enhance prediction.

\begin{figure}
\centering
    \includegraphics[width=\linewidth]{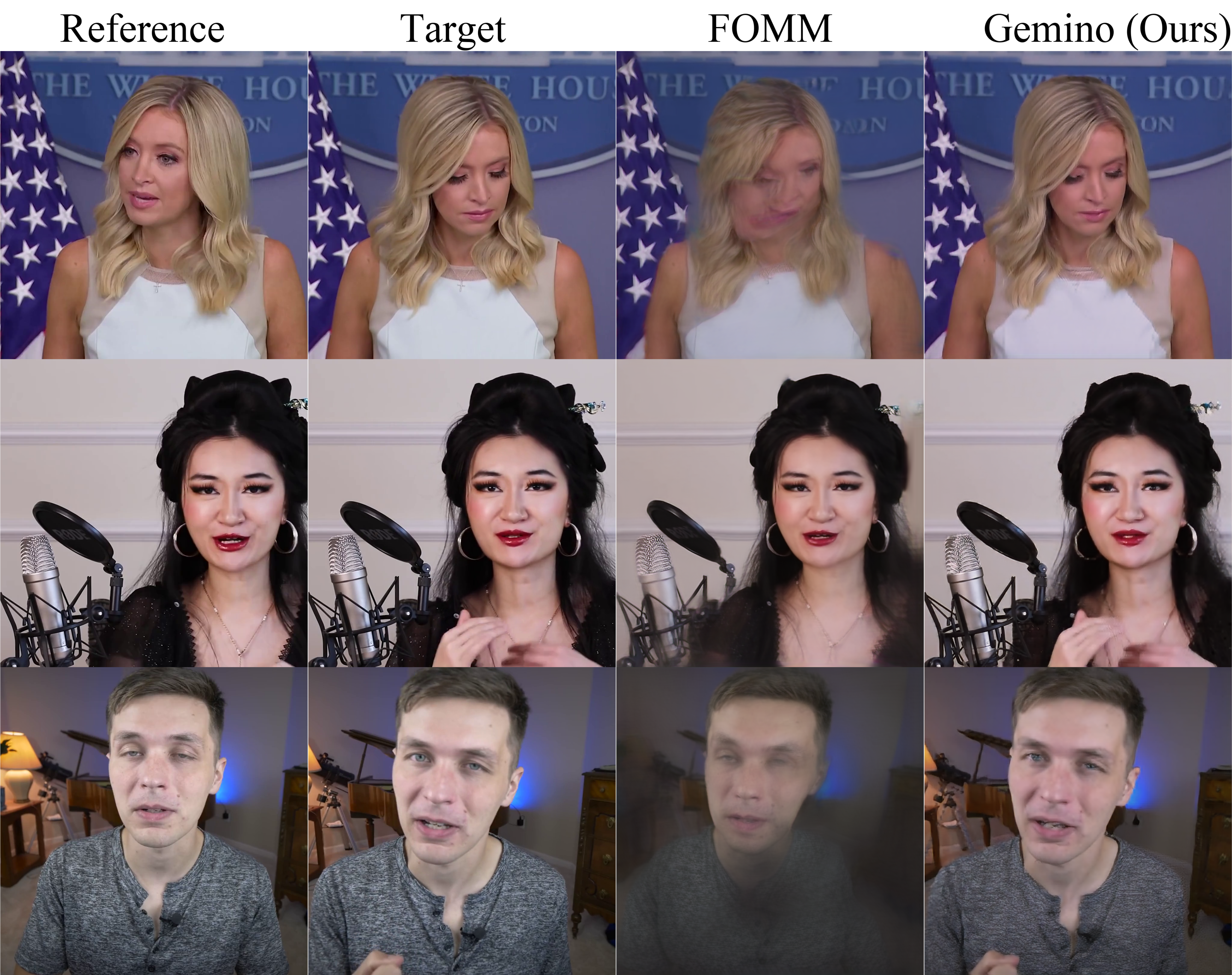}
\caption{\small Failure cases in the FOMM's~\cite{fom} reconstruction when the reference and target differ. FOMM misses the hand (row 2) because the reference frame does not contain it. In rows 1 and 3, FOMM only produces a blurry outline of the face (and torso). 
\TheSystem improves on these reconstructions by utilizing the downsampled target frame in its architecture and capturing the difference between the reference and the target better.}
\vspace{-15pt}
\label{fig:fom failures}
\end{figure}

\NewPara{Challenges for neural face image synthesis.}
\label{sec:motivation}
Neural synthesis approaches and specifically, \kp-based models fall short in a number of ways that make them impractical in a \vcing setting. These models operate similarly to the model described in \Fig{model structure} but do not transmit or use the downsampled target frame. They extract \kps from the downsampled target frame, and transmit those instead. This choice causes major reconstruction failures when the reference and target frames are not close. \Fig{fom failures} shows the reconstruction produced by the First-Order-Motion Model (FOMM)~\cite{fom}, a \kp-based model, on 1024$\times$1024 frames.
We focus on the FOMM as a representative \kp-based model, but these limitations extend to other such models.
The FOMM only produces blurry outlines of the faces in rows 1 and 3 where the reference and target differ in orientation and zoom level respectively.
In row 2, the FOMM misses the arm altogether because it was not present in the reference frame and warping alone cannot convey the arm's presence.
Such failures occur because \kps are limited in their representation of differences across frames, and most warping fields cannot be modeled as small deformations around each \kp. Poor prediction quality in the event of such movements in video calls seriously disrupts the user experience.

Secondly, even in regions without much movement between the reference and the target frames, current approaches do not have good fidelity to high-frequency details. In row 2 of \Fig{fom failures}, the microphone does not move much between the reference and the target, but possesses a lot of high-frequency detail in its grille and stand. Yet, the FOMM has a poor reconstruction of that area. In row 1 of \Fig{fom failures}, it misses even those details in the hair that are similar between the reference and the target.
This issue becomes more pronounced at higher resolutions where more high-frequency content is present in each frame, and where the human eye is more sensitive to missing details.

\begin{figure*}
\centering
\includegraphics[width=0.83\linewidth]{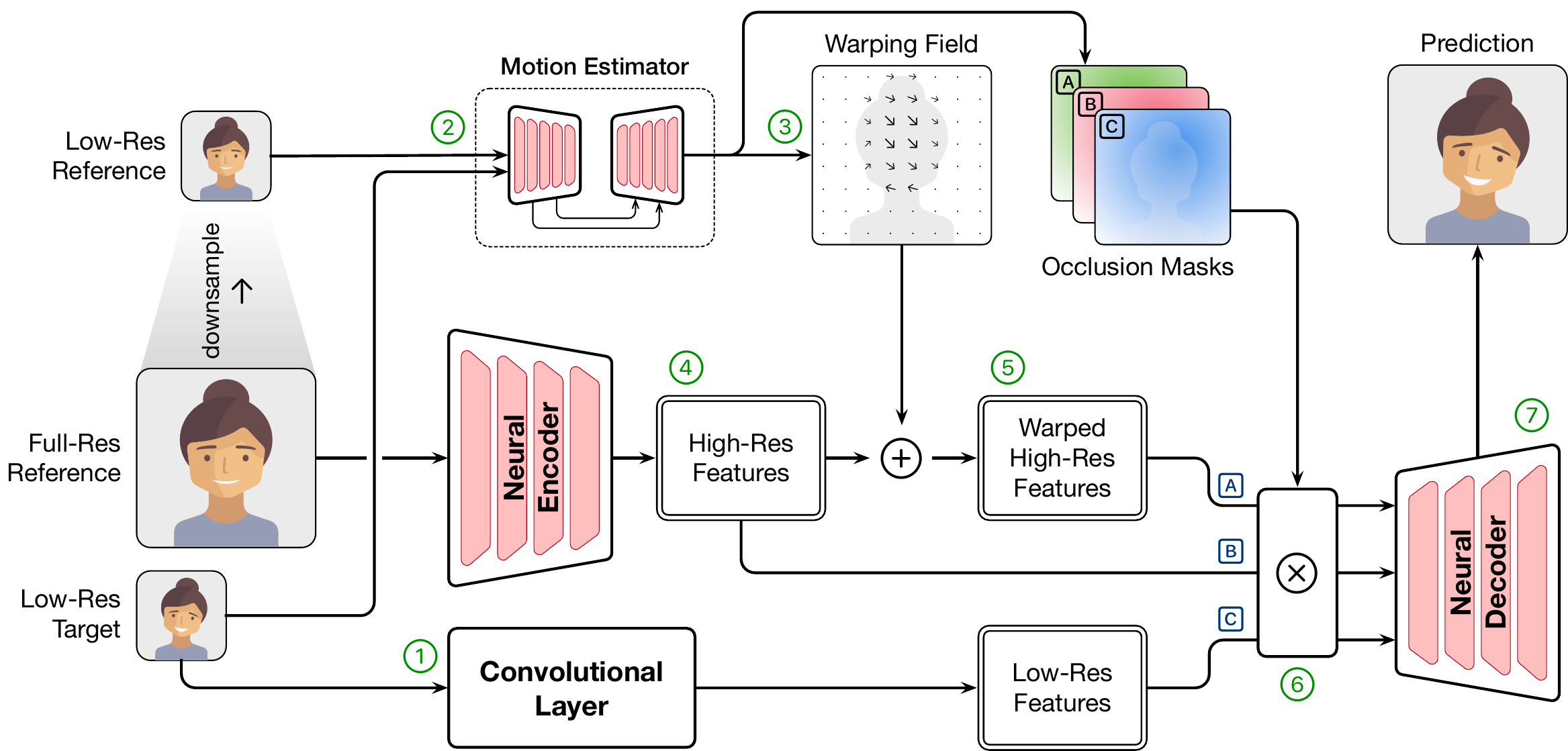}
\caption{\small \TheSystem's high-frequency-conditional super-resolution model at the receiving client. The model first obtains features from the low-resolution target. It combines the low-resolution reference and target frames to produce a warping field based on the motion between them. The warping field is applied on encoded features from a full-res reference frame. The low-resolution and full-resolution features are jointly decoded by the neural decoder to produce the prediction at the receiver. } 
\label{fig:detailed model diagram}
\vspace{-8pt}
\end{figure*}

Lastly, many existing approaches~\cite{fbfom,fom,fom_multiple_views} are evaluated on 256$\times$256 images~\cite{voxceleb}. 
However, typical video conferences are at least 720p, especially in the full-speaker view. While the convolutions in these models allow them to scale to larger resolutions with the same kernels, the inference times often exceed the real-time deadline (33ms per frame for 30 frames-per-second) at high resolutions. 
We observe this in \Tab{prediction_diff_resolution} which shows the inference time of the FOMM~\cite{fom} on different GPU systems at different resolutions. 
This is unsurprising: running the FOMM unmodified at different resolutions performs the same number of operations per-pixel on 1024$\times$1024 and 256$\times$256 frames even though each pixel represents a smaller region of interest in the former case. With 16$\times$ more input pixels, the model's encoded features are also larger. 
This suggests that we need to redesign neural codecs to operate at higher resolutions without significant compute overheads, particularly as we move towards 4K videos.
\section{System Design}
\label{sec:design}


\subsection{Overview}
\label{sec:design overview}
Our goal is to design a robust neural compression system that reconstructs both low and high frequency content in the target frame with high fidelity even under extreme motion and variance across target frames. Ideally, such a system should also be able to operate at high resolution without significant compute overheads. Our prototype operates at 1024$\times$1024 resolution.

Our key insight is that we need to go beyond the limited representation of movement that \kps or facial landmarks alone provide. Fortunately, modern video codecs are very efficient in compressing low-resolution (LR) video frames.
For instance, VPX compresses 128$\times$128 resolution  frames (8$\times$ downsampled in each dimension from our target resolution) using only \SI{15}{Kbps}. Motivated by this observation and the fact that downsampled frames possess much more semantic information than \kps, we design a model that performs \emph{super-resolution conditioned on high-frequency textures} from a full-resolution reference frame.
More specifically, our model resolves a downsampled frame
at the receiver to its full resolution, aided by features derived from a full-resolution reference frame.

\begin{table}[!t] 
    \centering
    \small
    \resizebox{\linewidth}{!}{
    \begin{tabular}{lccccc}
    \toprule
    & \multicolumn{3}{c}{\textbf{First-Order-Motion Model}} & \textbf{\TheSystem} & \textbf{\TheSystem (Opt.)}\\
    \cmidrule(lr){2-4} \cmidrule(lr){5-5} \cmidrule(lr){6-6}
    GPU & 256$\times$256 & 512$\times$512 & 1024$\times$1024 & $1024\times1024$ & $1024\times1024$\\
        \cmidrule(lr){1-1} \cmidrule(lr){2-6}
        Titan X & 17.35 ms & 50.13 ms & 187.30 ms & 68.08ms & 26.81ms\\
 V100 & 12.48 ms & 33.11 ms & 117.27 ms & 41.23ms & 17.42ms\\
 A100 & 8.19 ms & 12.76 ms & 32.96 ms & 17.66ms  & 13.89ms \\
    \bottomrule
    \end{tabular}}
    \caption{Inference time of the First-Order-Motion Model~\cite{fom} at different resolutions on different GPU systems. Prediction time increases with resolution, making it hard to reconstruct using the FOMM at 1024$\times$1024 in real-time. The \ms architecture (third column) and further optimizations (fourth column) allow \TheSystem to achieve real-time inference at 1024$\times$1024 on NVIDIA V100, A100, and Titan X systems.}
    \label{tab:prediction_diff_resolution}
    \vspace{-20pt}
\end{table}

\subsection{Model Architecture}
\label{sec:conditional SR}
\Fig{detailed model diagram} describes \TheSystem's model architecture running at the receiver of a \vcing session. We assume that the receiver has access to a single \emph{high-resolution reference frame}, capturing what the speaker looks like in that particular session. It also receives a stream of \emph{low-resolution target frames} that are compressed by a traditional video codec. \circled{1}~The receiver first takes the decompressed low-resolution (LR) target that it receives from the sender over the network and runs it through a single convolutional layer to produce LR features. \circled{2}~Next, the reference frame is downsampled and supplied, along with the (decompressed) LR target, to a motion estimation module that consists of a UNet~\cite{unet}. \circled{3}~Since the two frames typically differ in their axes alignment, the motion estimator uses the \kps to output a ``warping field'' or a transformation to move the reference frame features into the target frame's coordinates. 
\circled{4}~Meanwhile, the full-resolution reference frame is fed through a series of convolutional layers that encode it to extract high-resolution (HR) features.
\circled{5}~The model applies the transformation produced by the motion estimator to the HR features to produce a set of warped (or rotated) HR features in the target frame's coordinate system. 

Once the warped and encoded features from both the LR and HR pipelines are obtained, the model combines them before decoding the features to synthesize the target image. Specifically, \circled{6}~the model uses three inputs to produce the final reconstruction: \sfboxed{A}~the warped HR features, \sfboxed{B}~the HR features prior to warping, and \sfboxed{C}~the LR features.  Each of these inputs on a feature-by-feature basis with three occlusion masks of the same dimension, produced by the motion estimator. These occlusion masks sum up to 1 and describe how to weigh different regions of the input features when reconstructing each region of target frame. For instance, mask \sfboxed{A} maps to parts of the face or body that have moved in the reference and need to be regenerated in the target pose, mask \sfboxed{B} maps to the regions that don't move between the reference and target (\eg background), while mask \sfboxed{C} maps to new regions (\eg hands) that are only visible in the LR frame. \circled{7}~Once multiplied with their respective masks, the combined input features are fed through convolutional layers in the decoding blocks that upsample them back into the target resolution.  A more detailed description of the model with figures can be found in \App{model details}.


\NewPara{Why is super-resolution insufficient?} A natural question is whether the high-resolution information from the HR reference frame is essential or if we can simply perform super-resolution on the LR input frame through a series of upsampling blocks. While the downsampled LR frame is great
at conveying low-frequency details from the target frame, it possesses little high-frequency information. 
This often manifests in the predicted frames as a blur or lack of detail in clothing, hair, or facial features such as teeth or eye details. To synthesize frames that are faithful to both the low and high-frequency content in the target, it is important to combine super-resolution from the LR frame with features extracted from the HR reference frame~(\Sec{sec:eval_model}). The former ensures that we good low-frequency fidelity, while the latter handles the high-frequency details. The last column of \Fig{fom failures} shows how much the reconstruction improves with our high-frequency-conditional super-resolution approach. Specifically, we are able to capture the hand movement in row 2, the head movement in row 1, and the large motion in row 3. \TheSystem also produces a sharper reconstruction of facial features and the high-frequency content in the grille in row 2.

\begin{table}[t] 
    \centering
    \small
    \resizebox{0.65\linewidth}{!}{
    \begin{tabular}{rrr}
    \toprule
        Bitrate Regime & Resolution & Codec\\
        \midrule
        $< $\SI{30}{Kbps} & 128$\times$128 & VP8  \\
        \SI{30}{Kbps} -- \SI{75}{Kbps} & 256$\times$256 & VP8\\
        \SI{75}{Kbps} -- \SI{200}{Kbps} & 512$\times$512 & VP9\\
        $> $\SI{200}{Kbps} &   1024$\times$1024 & VP9 \\
    \bottomrule
    \end{tabular}}
    \caption{Mapping between desired bitrate regime and chosen resolution in \TheSystem.}
    \label{tab:resolution_bitrate_regime}
    \vspace{-10pt}
\end{table}

\subsection{Optimizations To Improve Fidelity}
\label{sec:system optimization fidelity}
\NewPara{Codec-in-the-loop training.}
Our design decision to transmit low-resolution frames instead of \kps is motivated by modern codecs' ability~\cite{vp8, vp9, av1} to efficiently compress videos at lower resolutions. Downsampled frames can be compressed at a wide range of bitrates depending on the resolution and quantization level. However, the LR video resolution determines the required upsampling factor. Note that different upsampling factors normally require different variants of the deep neural model (\eg different number of upsampling stages, different feature sizes, \etc). This means that for each target bitrate, we first need to choose a LR video resolution, and then train a model to reconstruct high-res frames from that resolution. We first create a reverse map from bitrate ranges to resolutions by profiling VPX codecs to identify the bitrate regime that can be achieved at each resolution. 
We use this map to select an appropriate resolution and codec for the LR video stream at each target bitrate. \Tab{resolution_bitrate_regime} shows the resolution and codec \TheSystem uses in each bitrate range.  Once a resolution is chosen, we train the model to upsample \emph{decompressed LR frames} (from the VPX codec) of that resolution to the desired output resolution. This results in separate models for each target bitrate regime, each optimized to learn the nature of frames (and artifacts) produced by the codec at that particular resolution and bitrate. 
Since each such model requires 30 hours of training time on a single A100 GPU, we show that it is prudent
to downsample to the highest resolution compressible at a particular bitrate, and use the model trained with a target bitrate at the lowest end of the achievable bitrate range  for that resolution  (\Sec{sec:eval_config}). 

\begin{figure}
    \centering
    \includegraphics[width=\linewidth]{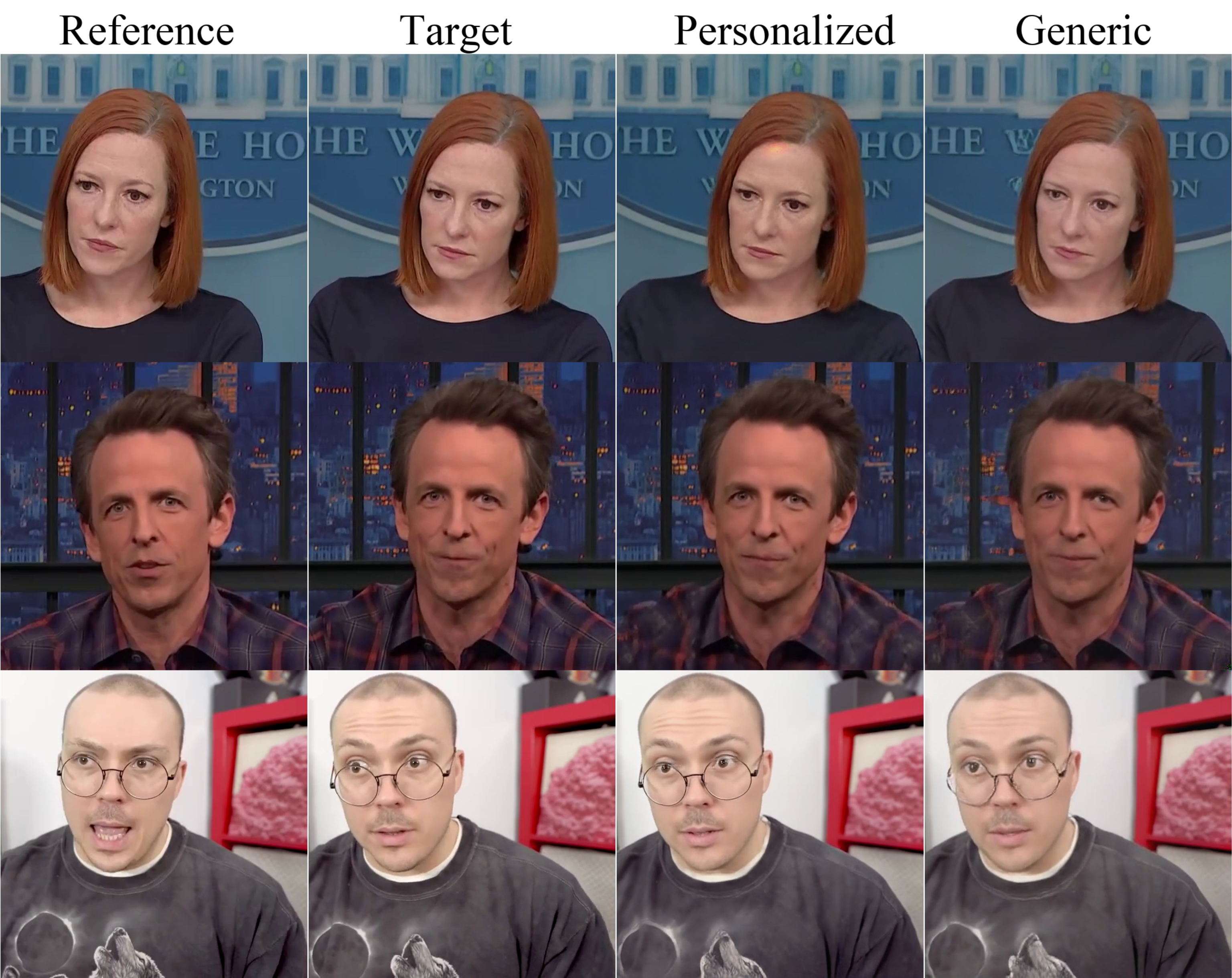} 
\caption{\small Performance of a model trained on a generic corpus of 512$\times$512 videos compared to a personalized model fine-tuned on the specific person in the video. The personalized model better captures the eye gaze, dimples, and rim of the glasses.}
\vspace{-14pt}
\label{fig:personalization vs generic 512}
\end{figure} 
\NewPara{Personalization.} To improve the fidelity of our reconstructions, 
we explore training \TheSystem in two ways: \emph{personalized} to each individual, and \emph{generic}.
To train the personalized model, we separate videos of each person into non-overlapping test and train data; the model is not exposed to the test videos, but learns a person's facial features from training videos of that specific person. 
The generic model is trained on a corpus of videos consisting of many different people.
\Fig{personalization vs generic 512} visually compares these two approaches. The personalized model is better at reconstructing specific high-frequency details of the person, e.g., eye gaze (row 1), dimples (row 2), and the rim of the glasses (row 3), compared to the generic model. 
We envision the model in neural \vcing systems to be personalized to each individual
using a few hours of their video calls, and cached at the systems of receivers who frequently
 converse with them. An unoptimized (full-precision) checkpoint of the model's 82M parameters is about a GB in size. However, we show that the model can be compressed in \Sec{sec:eval_config}. 
 

\subsection{Reducing Computational Overheads}
\label{sec:system optimization compute}
\NewPara{Multi-scale architecture.}
As we scale up the desired output resolution,
it is crucial to perform fewer operations per pixel to reduce the compute overheads. We carefully examine different parts of the model in \Fig{detailed model diagram}, and design a \ms architecture where we separate those modules that require fine-grained detail from those that only need coarse-grained information. Specifically, the motion estimation module is responsible for obtaining high-level motion information 
from the model. Even as the input resolution is increased, this coarse information can be inferred from low-resolution frames which retain low-frequency details. 
Consequently, in \TheSystem, we operate the motion estimation module on low-resolution reference and target frames (64$\times$64). In contrast, the neural encoder and decoders are responsible for reconstructing high-frequency content and require  fine-grained details from the high-resolution reference frame. 

To further reduce computational overheads, we adjust the number of downsampling and upsampling blocks in the neural encoder and decoder based on the input resolution, keeping the size of the intermediate bottleneck tensor manageable. For example, for 1024$\times$1024 input, we use 4 down/up-sampling blocks in the encoder and decoder (\Fig{detailed model diagram}). So as to not lose information through the bottleneck, 
we equip the first two blocks 
with skip connections~\cite{unet} that directly provide encoded features to the decoder.
As seen in the third column of \Tab{prediction_diff_resolution}, our \ms architecture reduces the reconstruction time of the model
by 2.75$\times$ on older Titan X and NVIDIA V100 GPUs, while allowing us to easily meet the real-time deadline on an NVIDIA A100 GPU. 

\NewPara{Model Optimizations.}
While the \ms architecture achieves real-time inference on A100 GPU system, it still fails to meet the 30 fps requirement on older systems such as Titan X. To enable this, we optimize our model further through a suite of techniques aimed at lowering the compute overheads of large neural networks.
Specifically, we focus on reducing the number of operations or Multiply-Add Cumulations (MACs) involved in the expensive decoding layers which are run on every frame. While a reduction in MACs does not always translate to reduction in latency on the targeted platform~\cite{eccv2018-yang}, it acts as a reasonable proxy for our case. We first replace the regular convolutional blocks in \TheSystem with depthwise separable convolutions (DSC) followed by a pointwise convolution~\cite{howard2017mobilenets, Wofk2019FastDepthFM}, a commonly used architecture for mobile inference. 

To decrease the model MACs even further without impacting its fidelity, we run NetAdapt~\cite{eccv2018-yang}, a neural-architecture search algorithm that iteratively prunes the number of channels layer-by-layer until the compute overheads fall below a target threshold. NetAdapt can directly optimize for latency on different platforms, but we instead use it to optimize the MACs to reduce the overheads of running separate architecture searches for each platform. NetAdapt achieves its MAC reduction in iterations, each of which targets a small decrease (\eg 3\%) towards its target. In each iteration, NetAdapt prunes the layer of the model which decreases its MACs with the least loss in accuracy after ``short-term finetuning'' on a small subset of the data. The pruned model acts as the starting point for the next iteration, and this process continues until the model has been shrunk sufficiently. Finally, the model is ``long-term fine-tuned'' on the entire dataset to recover its lost accuracy. 

Since \TheSystem uses personalized models, we adapt NetAdapt's logic to suit \TheSystem's training paradigm. We first apply NetAdapt on the generic \TheSystem model with short-term finetuning on the generic dataset. This is because neural architecture search is expensive to run, and we observe that the pruned architecture at the end of NetAdapt is the same for generic and personalized models.
We finally long-term finetune the shrunk model in a personalized manner to better recover its accuracy (\Sec{sec:eval_config}). These optimizations allow us to run inference on the Titan X GPU in \SI{27}{ms} with barely any loss in visual quality (last column of \Tab{prediction_diff_resolution}) when upsampling 128$\times$128 frames at \SI{15}{Kbps} to 1024$\times$1024. Given that \vcing applications tolerate latencies of up to 200 ms (5--6 frames) in their jitter buffers~\cite{itu}, we believe that the additional delay from generating the received frame will be negligible. 

\subsection{Operational Flow}
\NewPara{Training Procedure.}
To train \TheSystem, we first obtain weights from a trained FOMM model on the entire VoxCeleb dataset~\cite{voxceleb} at 256$\times$256 resolution. We choose the appropriate training data for the specific person we want to train a \TheSystem model for, and train from scratch the additional downsampling and upsampling layers in the HR pipeline as well as all layers in the LR pipeline, while fine-tuning the rest of the model for 30 epochs. We repeat this procedure for different LR frame resolutions and target bitrate regimes mention in \Tab{resolution_bitrate_regime}, and different people. In parallel, using the same procedure, we also train a generic version of the model on a larger corpus of people. Both models are replaced with depthwise-separable convolutions and optimized using NetAdapt~\cite{eccv2018-yang} to produce the final model.

\NewPara{Inference Routine.} Once versions of the model have been trained and optimized for different LR resolutions and target bitrates, we simply use the appropriate model for the current target bitrate regime and the person on the video call. The sender and the receiver pre-negotiate the reference frame at the beginning of the video call. This model performs inference on a frame-by-frame basis in real-time to synthesize the video stream at the receiver. We detail our prototype implementation and WebRTC pipeline further in \Sec{sec:impl}.

\section{Implementation}
\label{sec:impl}

\begin{figure}
    \includegraphics[width=\linewidth]{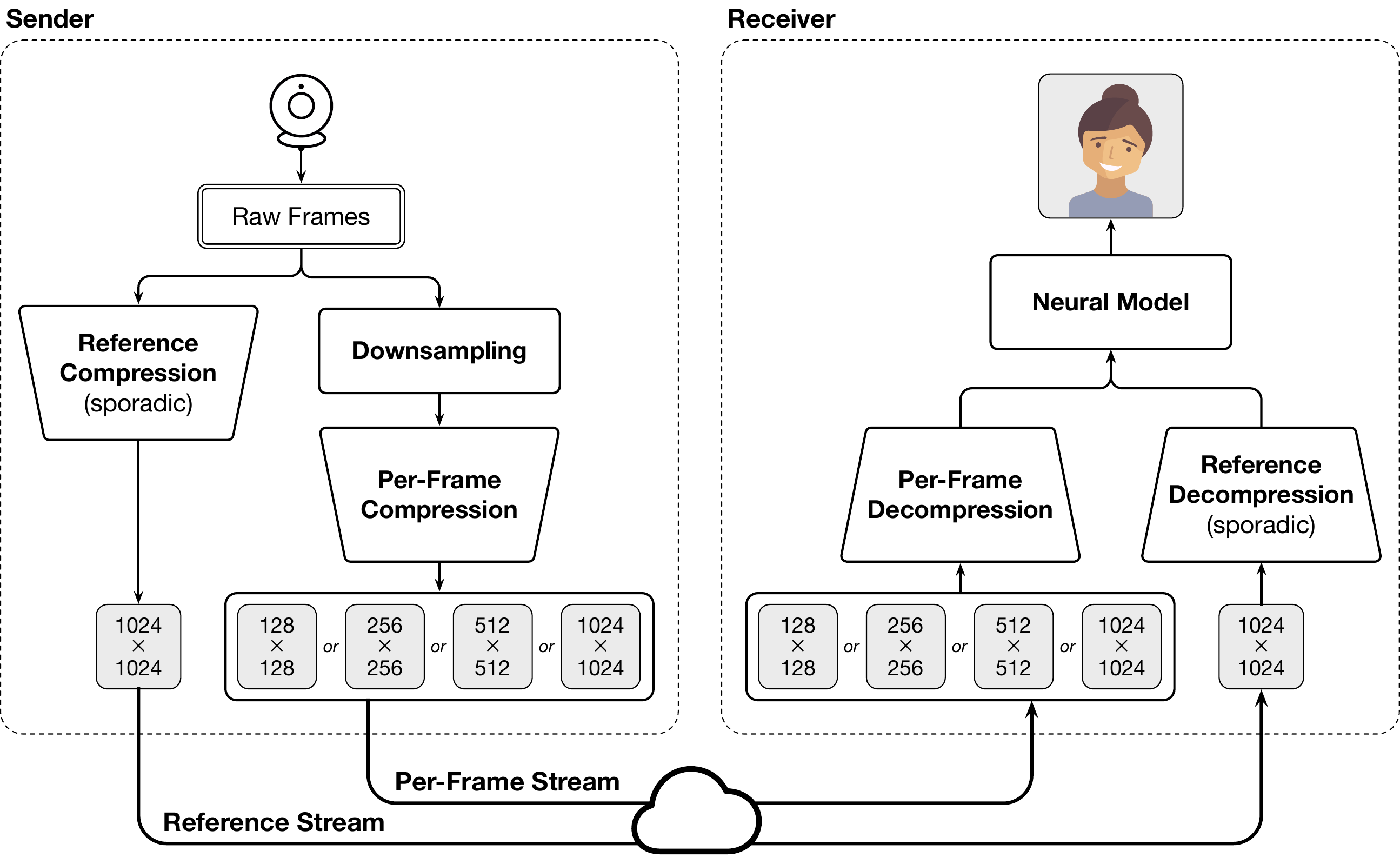}
    \centering 
      \caption{\small Neural video compression pipeline atop WebRTC~\cite{webrtc}. We use two RTP streams: A sparse \REF stream that sporadically sends high-resolution reference frames, and a \perframe (\PF) stream that is used on every frame. The \PF stream sends downsampled frames of the highest resolution that the current bandwidth can support, and thus has separate VP8 compression modules for each resolution. The receiver decompresses the downsampled frames, and supplies them, along with the latest reference frame, to the neural network that reconstructs the target video. If bandwidth is high enough, the \PF stream is used for full-resolution VP8 frames without synthesis.}
      \label{fig:software stack}
      \vspace{-8pt}
 \end{figure}

\NewPara{Basic WebRTC Pipeline.} Our neural \vcing solution uses WebRTC~\cite{webrtc}, an open-source framework that enables video and audio conferencing atop the real-time transport protocol (RTP)~\cite{schulzrinne1996rtp}. Since we perform neural frame synthesis, we use a Python implementation of WebRTC called \aiortc~\cite{aiortc} that allows easy interfacing with PyTorch. Aiortc handles the initial signaling and the peer-to-peer connection setup. A typical video call has two streams (video and audio) that are multiplexed onto a single connection. The sender extracts raw frames from the display, and compresses the video and audio components separately using standard codecs like VPX~\cite{vp8, vp9}, H.264/5~\cite{h264, h265}, Opus~\cite{opus}, \etc. The receiver decompresses the received data in both streams before 
synchronizing them and displaying each frame to the client. 


\NewPara{New Streams.} 
We extend the standard WebRTC stack to use two distinct streams for video: a \emph{\perframe stream} (\PF stream) that transmits downsampled video (e.g. 64$\times$64 frames) {\em on every frame}, and a \REF stream that transmits occasional but high-resolution {\em reference} frames that improve the synthesis fidelity. We anticipate using the \REF stream extremely sparsely. For instance, in our implementation, we use the first frame of the video as the \emph{only} reference frame. However, more reference frames may help recover high-frequency fidelity as it worsens when the reference and target frames drift apart. But, most low-frequency changes between the reference and the target can be communicated simply through the downsampled target in the \PF stream\footnote{We observe that sending reference frames with any fixed frequency adds significant bandwidth overheads. So, we only use a single reference frame in our evaluations. We leave an investigation of mechanisms to detect the need for a new reference frame (speaker moves significantly, high-frequency content or background changes) to future work.}. 
The receiver uses the per-frame information in the \PF stream, with the reference information, to synthesize each high-resolution frame. \Fig{software stack} illustrates the expanded WebRTC architecture to accommodate the \TheSystem design. 

The \PF stream is implemented as a new RTP-enabled stream on the same peer connection between the sender and the receiver.  
We downsample each input frame to the desired resolution at the sender and compress it using the appropriate VPX codec. The frame is decompressed at the receiver. The bitrate achieved is controlled by supplying a target bitrate to VPX. Our \PF stream can support full-resolution video that is typical in most \vcing applications, while also supporting a range of lower resolutions for the model to upsample from. To enable this flexibility, we design the \PF stream to have multiple VPX encoder-decoder pairs, one for each resolution that it operates at. When the sender transmits a frame, it chooses an appropriate resolution and codec based on the target bitrate, and compresses the video at that resolution and target bitrate. The resolution information is embedded in the payload of the RTP packet carrying the frame data. When the receiver receives each RTP packet, it infers the resolution and sends it to the VPX decoder for that resolution. Once decompressed, the low-resolution frame is upsampled by \TheSystem to the appropriate full-resolution frame. If the \PF stream consists of 1024$\times$1024 frames, \TheSystem falls back onto the regular codec and stops using the \REF stream. The \REF stream is repurposed from the existing video stream.


\NewPara{Model Wrapper}.
To enable neural frame synthesis, 
we define a wrapper that allows the \aiortc pipeline to interface with the model. We reuse most of the pipeline from frame read at the sender to display at the receiver, except for introducing a downsampling module right after frame read, and a prediction function right before frame display. The wrapper is structured to perform format conversions and data movement from the AudioVisual~\cite{pyav} frames on the CPU that \aiortc needs, to the PyTorch~\cite{NEURIPS2019_9015} tensors on the GPU required by the model. We initialize models separately
for the sending and receiving clients. The wrapper also allows us to save (and periodically update) state at the sender and receiver which is useful for reducing the overheads from modules where we can reuse old computation (\eg run the encoder for high-resolution reference features only when the reference changes). 

\NewPara{Further Optimizations.} We optimize a number of other aspects of the \aiortc pipeline. 
For instance, we move data between the CPU and GPU multiple times in each step of the pipeline. 
Batching these operations is difficult when maintaining low latency on each frame. 
However, to minimize PCIe 
overheads from repetitive data movement, we use \emph{uint8} variables instead of \emph{float}. We also keep reference frames and their encoded features stored as model state on the GPU. We pipeline as many operations as possible by running \kp extraction, model reconstruction, and conversions between data formats in separate threads. 



\section{Evaluation}
\label{sec:eval}
We evaluate \TheSystem in a simulation environment and atop a WebRTC-based implementation. We describe our setup in \Sec{sec:eval_setup} and use it to compare existing baselines in \Sec{sec:eval_overall}. \Sec{sec:eval_model} motivates our model design, \Sec{sec:eval_config} discusses the impact of having the codec in our training process, and \Sec{sec:eval_adaptation} shows that \TheSystem closely matches a time-varying target bitrate. 
\subsection{Setup}
\label{sec:eval_setup}
\NewPara{Dataset.} 
Since most widely used datasets are of low-resolution videos~\cite{voxceleb,voxceleb2,maxine} and lack diversity in the extent of the torso or face-zoom level, we collected our own dataset comprising of videos of five Youtubers with publicly available HD (1920$\times$1080) videos.
For each Youtuber, we curate a set of 20 distinct videos or URLs 
that differ in clothing, hairstyle, accessories, or background. 
The 20 videos of each Youtuber are separated into 15 training videos and 5 test videos. 
For each video, we manually record and trim the segments that consist of talking individuals; 
we ignore parts that pan to news segments or different clips.
The segments are further split into 10s chunks to generate easily loadable videos for training, 
while the segments of the test video are combined to form a longer video.
We also
spatially crop each frame into our desired dimensions (typically 1024$\times$ 1024). 
based on the average location of the person
across all frames of the video. Note that 720p and 1024$\times$1024 frames have similar numbers of pixels. We strip the audio since our focus is on video synthesis.
\Tab{dataset_info} in \App{training} describes the details of the dataset. We do not own any of these videos, and we only use images of frames produced by our evaluation pipeline in this paper.
We use the 512$\times$512 dataset from NVIDIA~\cite{maxine} to train a generic model to illustrate the benefits of personalization. Our evaluation focuses on reconstructing a single front-facing person in a video call; \TheSystem can be extended to multiple speakers if there are application-level techniques to separate speakers into individual streams~\cite{meet-rooms,zoom-rooms}.

\NewPara{Model Details.}
The main model we evaluate is our high-frequency conditional super-resolution model that consists of an upsampling module that takes in features from a low-resolution (LR) frame, and upsamples it to 1024$\times$1024. To provide the high-frequency details, it uses two pathways consisting of warped and unwarped features from the high-resolution (HR) reference image (\Fig{detailed model diagram}). We use the first frame of the video as the \emph{sole reference} image for the entire test duration. The warping field is produced by a motion estimation network that uses the first-order approximation near 10 \kps~\cite{fom}. 
Our \ms architecture runs motion estimation always at 64$\times$64 irrespective of the input video resolution.
The neural encoder (for the HR features) and decoder (for both LR and HR features) consist of four down and upsample blocks. The discriminator operates at multiple scales and uses spectral normalization~\cite{spectralnorm} for stability. Layers of our model that are identical in dimensions to those from the FOMM are initialized from a public FOMM checkpoint trained on the VoxCeleb dataset~\cite{voxceleb}, and fine-tuned on a per-person basis. 
The remaining layers are randomly initialized and trained on a per-person basis over 30 epochs. We fine-tune the FOMM baseline also in the same personalized manner. We use Adam optimizer~\cite{kingma2014adam} to update the model weights with a learning rate of 0.0002, and first and second momentum decay rates of 0.5 and 0.999. We use equally weighted multi-scale VGG perceptual loss~\cite{johnson2016perceptual}, a feature-matching loss~\cite{wang2018video}, and a pixel-wise loss. We also use an adversarial loss~\cite{goodfellow2014generative} with one-tenth the weight of remaining losses. The \kps use an equivariance loss similar to the FOMM~\cite{fom}. We train our models to reconstruct from decompressed VPX frames corresponding to the low-resolution target frame so that the model learns to correct any artifacts produced by VPX. 


\NewPara{Evaluation Infrastructure.}
We evaluate our neural compression system in a simulation environment where frames are read from a video, downsampled (if needed) for the low-resolution \PF stream, compressed using VPX's chromium
codec~\cite{vp8_chromium}, and passed to the model (or other baselines) to synthesize the target frame. Note that the FOMM~\cite{fom} uses \kps and four ``Jacobian'' values around each \kp for producing its warping, and transmits them over the network. We design a new codec for the \kp data that achieves nearly lossless compression and a bitrate of about \SI{30}{Kbps}.
For VPX, we compress and decompress the full-resolution frame at different target bitrates, and measure the difference in visual quality between the output and the original frame. 

To obtain end-to-end latency measurements and to demonstrate \TheSystem's adaptability to different target bitrates, (\Sec{sec:eval_adaptation}), we use
our \aiortc~\cite{aiortc} implementation. 
A sending process reads video from a file frame-by-frame 
and transmits it to a receiving process that records each received frame. 
The two processes, running on the same server,
use the ICE signaling~\cite{ICE-connection}
mechanism to establish a peer-to-peer connection over a UNIX socket, which then supports video frame transmission using the Real-Transport Protocol (RTP). 
We timestamp each frame as it is sent and received, and save the sent and received frames in their uncompressed forms to compute latency and visual metrics.
We also log RTP packet sizes to compute the bitrate. 

\NewPara{Metrics.}
To quantify the aesthetics of the generated video, we use standard visual metrics such as PSNR (peak signal-to-noise ratio), SSIM (structural similarity index) in decibels~\cite{ssim}, and LPIPS (learned perceptual image patch similarity)~\cite{lpips}. For PSNR and SSIM, higher is better; while for LPIPS, lower is better. 
We observe that differences in LPIPS are more reflective
of how natural the synthesized frame feels and use that as our main comparison metric (~\Sec{sec:app:metrics}); we also show visual strips where appropriate.
We report the bitrate consumed to achieve a particular visual quality by measuring the total data transferred (size of compressed frames or RTP packet sizes) over the duration of the video, and dividing it by the duration itself.
To measure the end-to-end latency, we record the time at which the frame is read from the disk at the sender as well as the time at which prediction completes at the receiver. We report the difference between these two timestamps as our per-frame latency metric. 
We also report the inference time per frame when running the trained model in simulation; this does not capture the overheads of data conversion or movement in an end-to-end pipeline. 
This inference time needs to be $<33$ms to maintain a 30fps video call.
We run all our experiments for the entire duration of each test video in our dataset (\Tab{dataset_info}), and report the average over all frames for each metric.




\NewPara{Baselines.}
We obtain the bitrate for \emph{VP8}, the default codec in its Chromium settings~\cite{vp8_chromium} that 
comes with the \aiortc codebase. We also implement and evaluate VP9 in the same setup. 
To evaluate the benefits of using a neural approach to \vcing, particularly at lower bitrates, we compare a few different models: (1) \emph{FOMM}~\cite{fom}, a keypoint-based model for face animation,
(2) our approach, \TheSystem, 
(3) state-of-the-art super-resolution model based on \emph{SwinIR}~\cite{liang2021swinir}, and 
(4) \emph{bicubic upsampling}~\cite{bicubic} applied to the low-resolution VPX target frame.
All of the compared models generate 1024$\times$ 1024 frames except for the generic model that uses NVIDIA's 512$\times$512 corpus~\cite{maxine}.
 

\begin{figure}[!t]
\centering
\subfigure[\small Overall rate-distortion curve for all schemes]{
\includegraphics[width=\linewidth]{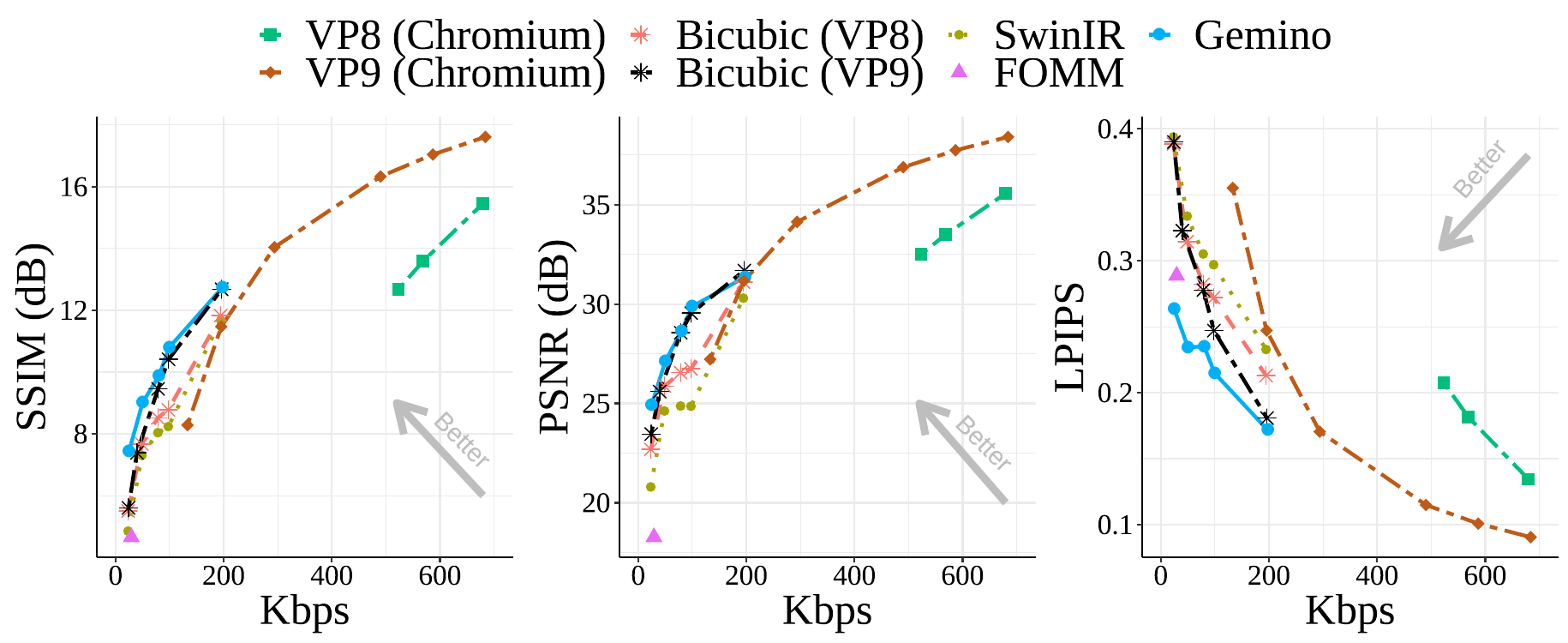}
\label{fig:main_comparison_full}
}
\subfigure[\small Rate-distortion curve in low-bitrate regimes.]{
\includegraphics[width=\linewidth]{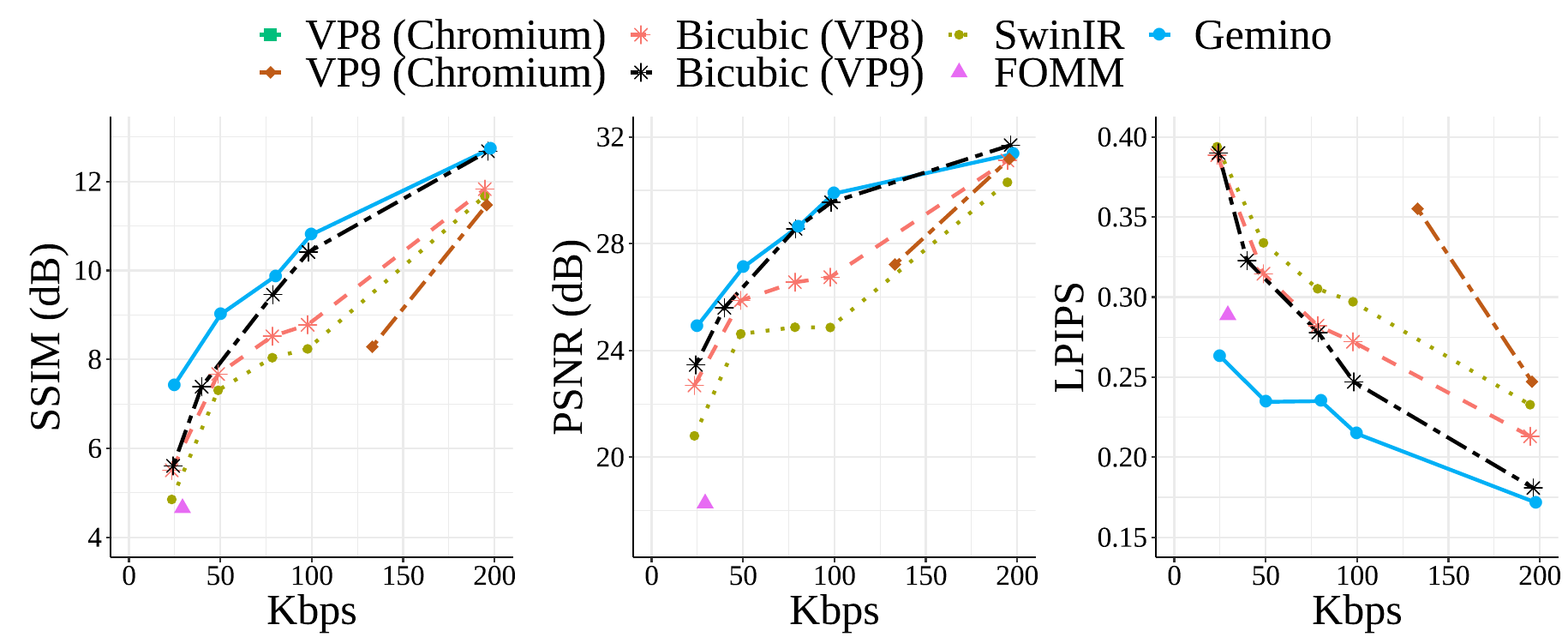}
\label{fig:main_comparison_low_bitrate}
}
\caption{\small Rate-distortion curve for \TheSystem compared with existing baselines. VP8 and VP9 require $\sim$5$\times$ and $\sim$3$\times$ the bitrate consumed by \TheSystem to achieve comparable LPIPS. At lower bitrates, \TheSystem outperforms other approaches that upsample low-resolution video frames. \TheSystem's benefits become prominent as the bitrate regime is lowered.}
\label{fig:main_comparison}
\vspace{-5pt}
\end{figure}

\begin{figure}[!t]
\centering
\includegraphics[width=\linewidth]{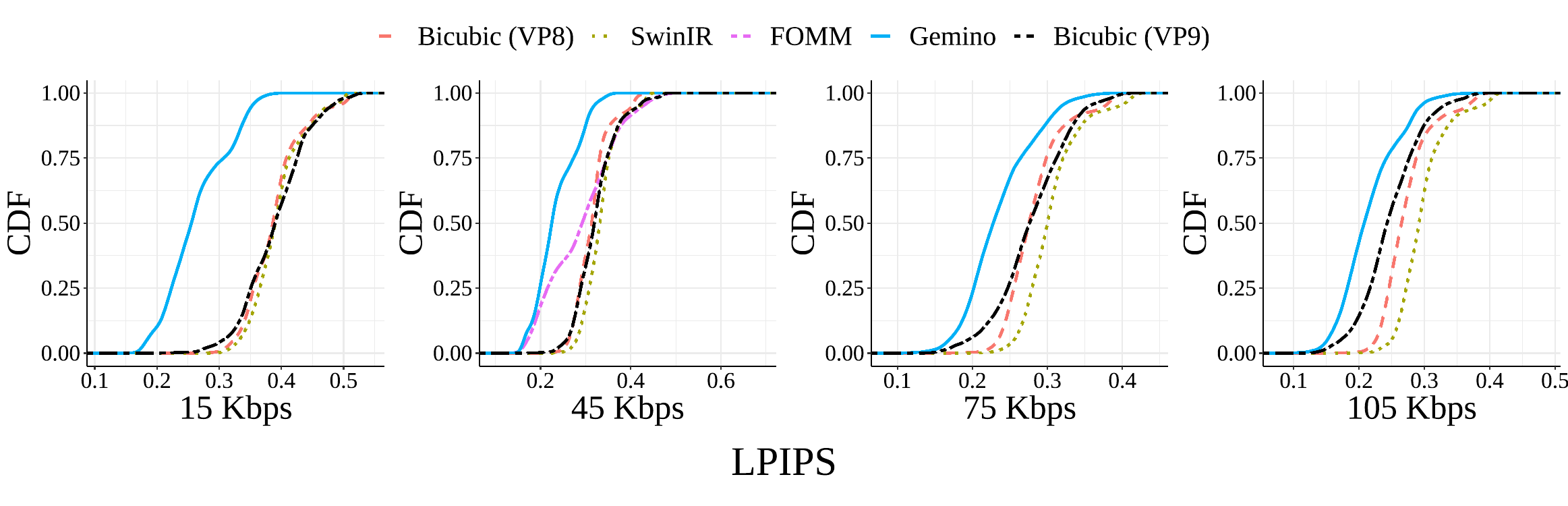}
\caption{\small CDF of reconstruction quality across all video frames as that shows that, as we move from higher bitrates to lower, the improvement from \TheSystem relative to Bicubic, particularly over VP9, becomes more pronounced.}
\label{fig:main_comparison_cdf}
\vspace{-12pt}
\end{figure}

\begin{figure*}[!t]
\centering
\begin{minipage}[b]{0.7\textwidth}
\includegraphics[width=\linewidth]{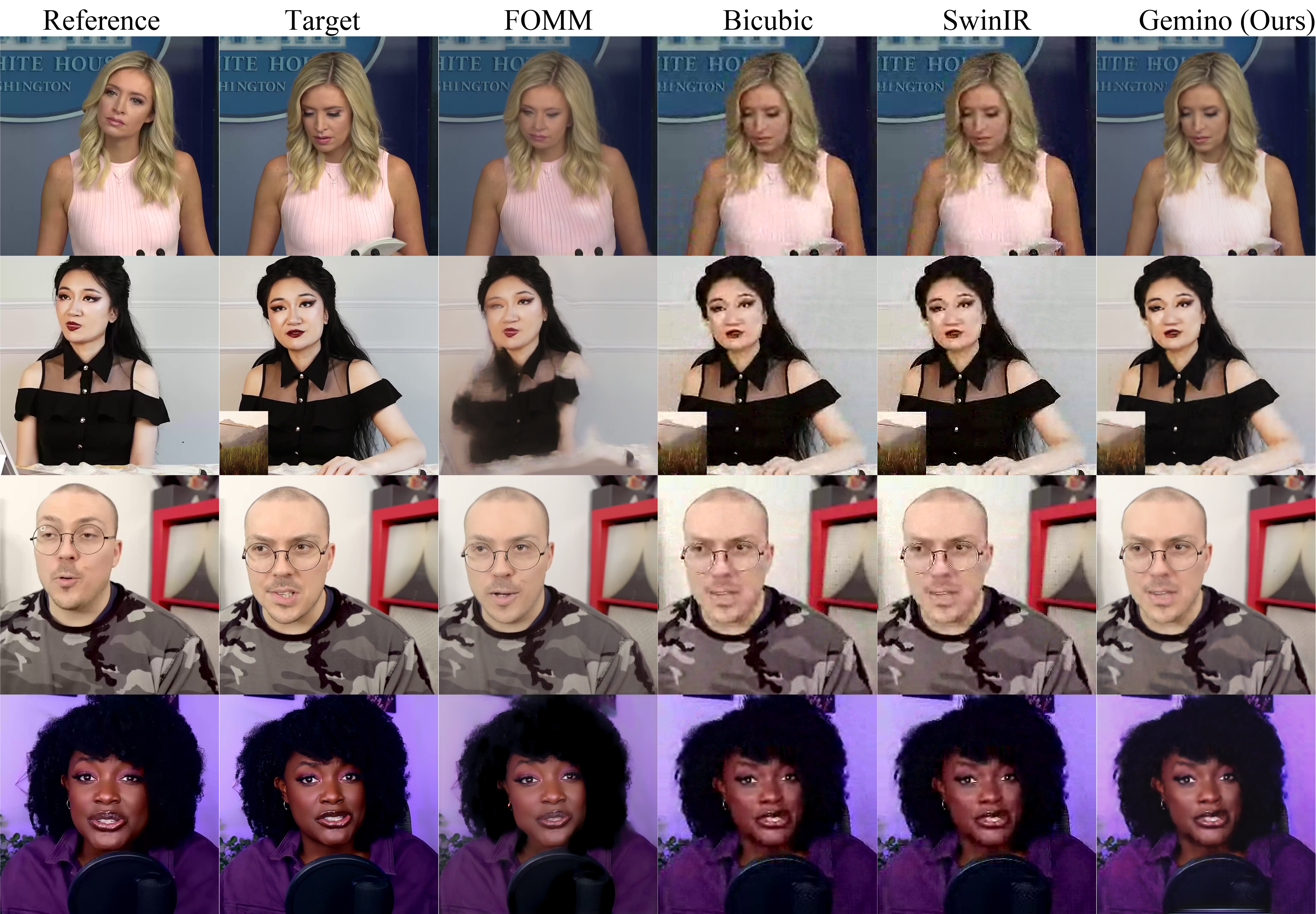}
\end{minipage}\hfill%
\begin{minipage}[b]{0.27\textwidth}%
\caption{\small Visual comparison across low-bitrate baselines. All but FOMM upsample a 256$\times$256 frame at 45 Kbps. \TheSystem's reconstructions contain a smoother output than the blocky artifacts observed with Bicubic and SwinIR. The FOMM completely fails when the reference and the target differ considerably in rows 1 and 2.}\vspace{-10pt}
\label{fig:main_comparison_strip}
\end{minipage}
\vspace{2ex}
\end{figure*}

\subsection{Overall Bitrate vs. Quality Tradeoff}
\label{sec:eval_overall}

To quantify the improvements of our neural compression system, we first compare \TheSystem with VP8 and VP9 in their chromium configuration~\cite{vp8_chromium}. \Fig{main_comparison} shows the rate-distortion curve for all schemes. For VPX, we alter the target bitrate alone for full-resolution (1024$\times$1024) frames in the \PF stream. For \TheSystem, bicubic, and SwinIR, we vary the resolution and target bitrate of the low-resolution (LR) frame in the \perframe (\PF) stream. For each point on the rate-distortion curve for \TheSystem, we train a personalized model to reconstruct full-resolution frames from LR frames encoded at the highest resolution supported by that target bitrate. We motivate using the codec in training and choosing the resolution in \Sec{sec:eval_config}. The \PF stream is compressed with VPX's Chromium settings at the target bitrate for its resolution. We configure \TheSystem to choose the base codec (VP8 or VP9) that has the better visual quality for a given target bitrate. We plot the resulting bitrate and visual metrics averaged across all 25 test videos' (5 speakers; 5 videos each) frames. \Fig{main_comparison_full} shows that VP8 operates in a different bitrate regime than all other schemes. VP9 improves the dynamic range to about \SI{130}{Kbps}, but still can't compress below that. Both VPX codecs operate on full-resolution frames, which cannot be compressed as efficiently as LR frames by the codec in its real-time mode. We observe that VP8 and VP9 consume \SI{523}{Kbps} and \SI{230}{Kbps} respectively to achieve an LPIPS of 0.21. In contrast, our approach is able to achieve the same visual quality with only \SI{105}{Kbps}, providing nearly 5$\times$ and 2.2$\times$ in improvement over VP8 and VP9 respectively. 



However, since our goal is to enable \vcing in bitrate regimes where current codecs fail, we focus on bitrates $<$ \SI{200}{Kbps} in \Fig{main_comparison_low_bitrate}. \TheSystem uses VP8 at \SI{15}{Kbps} and \SI{45}{Kbps} but VP9 at higher bitrates; VP8 and VP9 do not differ much at lower bitrates, but VP9 allows even 512$\times$512 to be compressed to \SI{75}{Kbps}. \Fig{main_comparison_low_bitrate} suggests that while LPIPS always shows considerable variation across schemes, small differences (1--2 dB) in PSNR and SSIM start manifesting only at lower bitrates. Specifically, \TheSystem achieves over \SI{1}{dB} better SSIM and PSNR, and 0.08 lesser LPIPS than Bicubic around \SI{50}{Kbps}. To map these metrics to improvements in visual quality, we show some snippets in \Fig{main_comparison_strip} and enlarged frames (with per-frame metrics) in \Sec{sec:app:metrics}. \Fig{main_comparison_strip} shows that compared to \TheSystem, reconstructions from bicubic have more block-based artifacts in the face, and the FOMM misses parts of the frame (row 4) or distorts the face. SwinIR, a super-resolution model, performs worse than bicubic. We suspect this is because SwinIR is not specifically trained on faces, and is also oblivious to artifacts from the codec that it encounters in our \vcing pipeline. To account for this, we evaluate a simple upsampling model personalized on faces in \Sec{sec:eval_model}. 

We also show CDFs of the visual quality in LPIPS across all frames in our corpus in \Fig{main_comparison_cdf} at \SI{15}{Kbps}, \SI{45}{Kbps}, \SI{75}{Kbps} and \SI{105}{Kbps}. The same CDFs for PSNR and SSIM can be found in \Sec{sec:app:low bitrate}. At each target bitrate, we use the largest resolution supported by the underlying video codec.
The CDFs show that \TheSystem's reconstructions are robust to variations across frames and orientation changes over the course of a video.
\TheSystem outperforms all other baselines across
all frames and metrics. Specifically, its synthesized frames at \SI{45}{Kbps} are better than FOMM 
by 0.05--0.1. It also outperforms Bicubic and SwinIR by 0.05 and 0.1 in LPIPS at the median and tail respectively. As we move from higher bitrates to lower, we observe that the improvement from \TheSystem relative to Bicubic, particularly over VP9, becomes more pronounced. While the gap in LPIPS at the median between Bicubic (VP9) and \TheSystem is less than 0.05 at \SI{105}{Kbps}, it increases to nearly 0.2 at \SI{15}{Kbps}. \TheSystem increases the end-to-end frame latency to $\sim$\SI{87}{ms}, from VPX's \SI{47}{ms}. However, our carefully pipelined operations are optimized to enable a throughput of 30 fps.\footnote{We ignore the delay due to the expensive conversion between NumPy arrays and Python's Audio Visual Frames in calculating latency because the the current PyAV library~\cite{pyav} is not optimized for conversion at higher resolutions.}

\subsection{Model Design}
\label{sec:eval_model}

\begin{figure}[!t]
\centering
\includegraphics[width=0.95\linewidth]{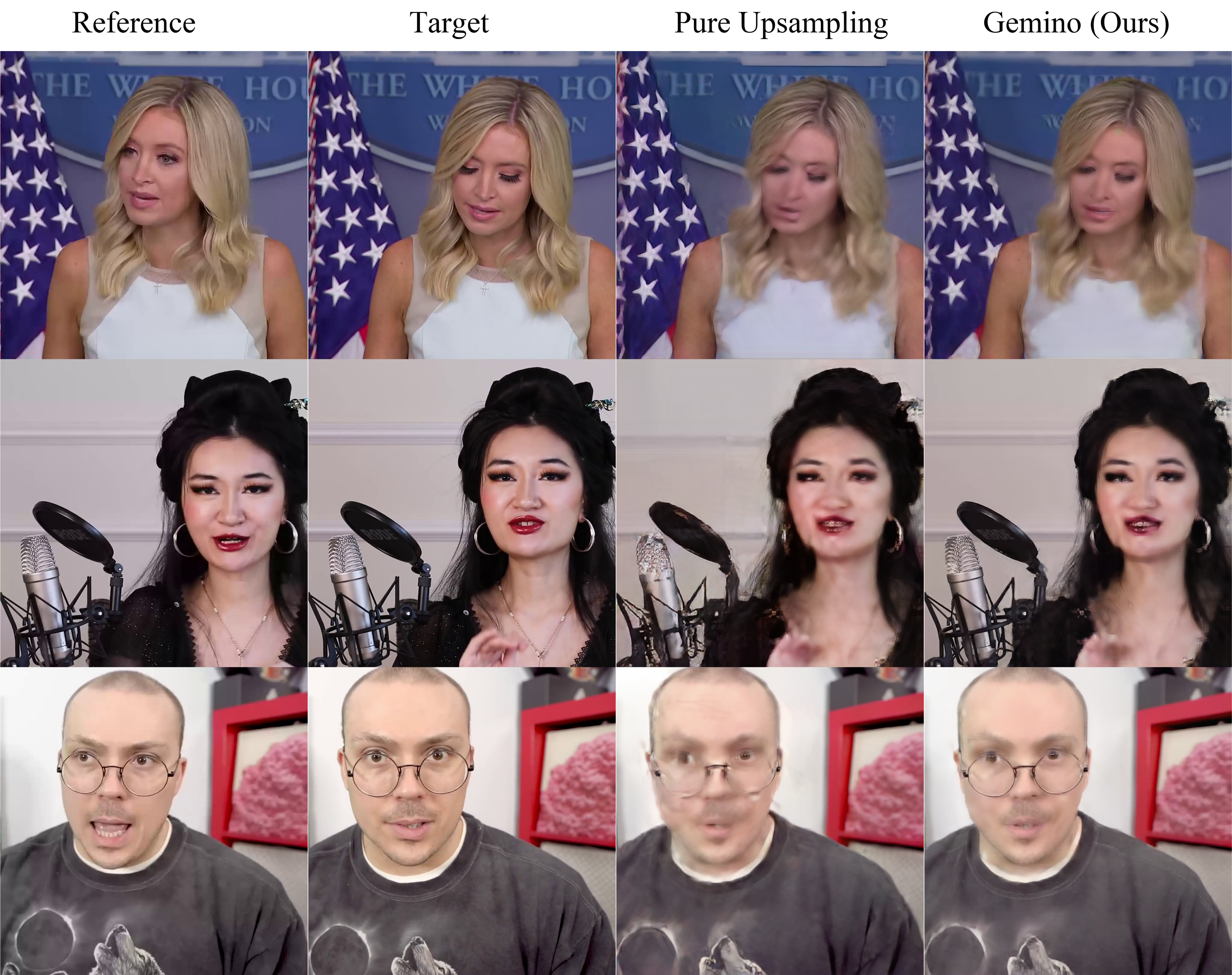}
\caption{\small Visual comparison across different model architectures. Compared to \TheSystem, Pure upsampling misses the high-frequency details of the microphone (row 2) and lends block-based artifacts on the face (row 3).}\vspace{-10pt}
\label{fig:model_ablation_strip}
\end{figure}

\begin{figure}
\centering
\includegraphics[width=0.77\linewidth]{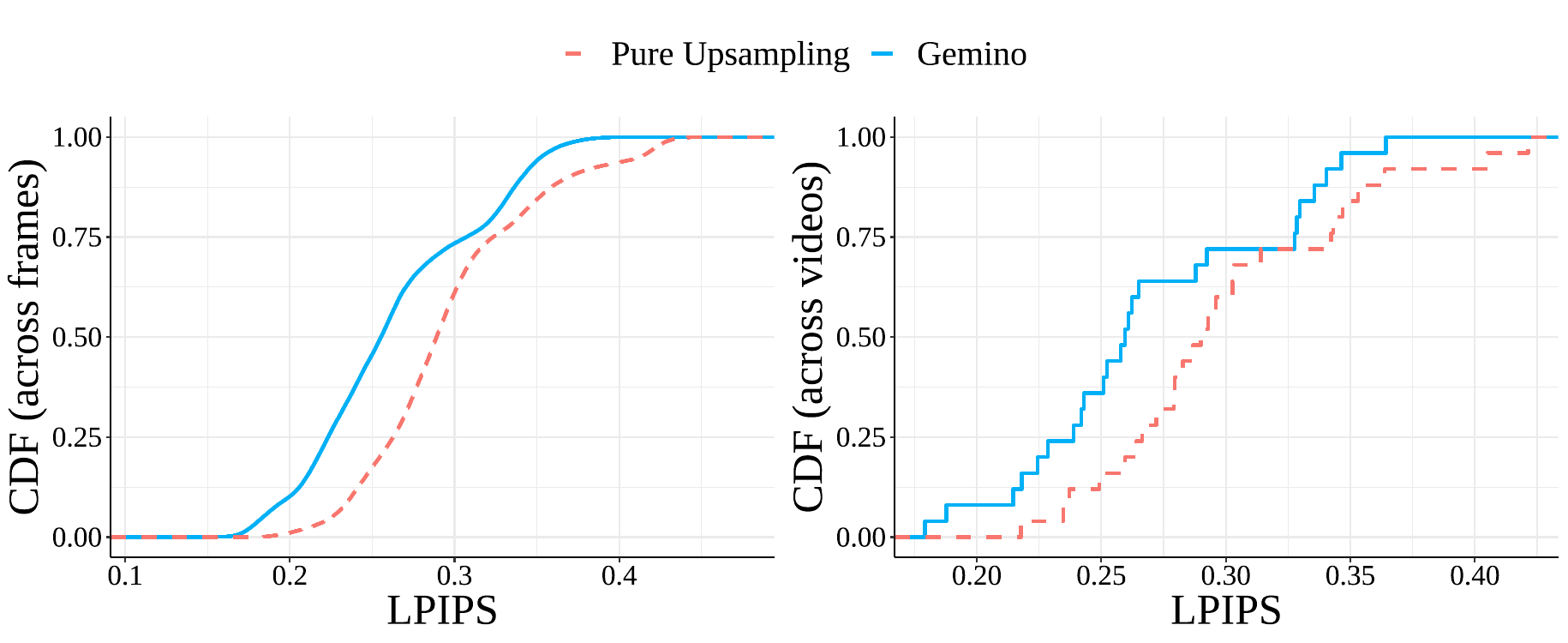}
\caption{\small CDF of reconstruction quality of different model architectures across frames and videos in our test corpus. \TheSystem outperforms pure upsampling by nearly 0.05 in LPIPS at the median of both frames and videos. It also outperforms the approach that relies on only warped HR, and the RGB-based warping across most  frames in the corpus.}
\label{fig:model_ablation_cdf}
\vspace{-10pt}
\end{figure}

\NewPara{Architectural Elements.} To understand the impact of our model design, we compare \TheSystem to a \emph{pure upsampling} approach from the LR target features with no pathways from an HR reference frame (only \sfboxed{C} in \Fig{detailed model diagram}) at decompressing VP8 128$\times$128 frames at \SI{15}{Kbps} to its full 1024$\times$1024 resolution. 
\Tab{model_architecture_comparison} reports the average visual quality while \Fig{model_ablation_cdf} shows a CDF of performance across all frames and videos. 
\TheSystem outperforms pure upsampling on average by \SI{0.4}{dB} on SSIM and 0.04 on LPIPS. This is also visible in rows 2 and 3 of \Fig{model_ablation_strip} where the pure upsampling approach misses the high-frequency details of the microphone grille and lends blocky artifacts to the face respectively. This manifests as a 0.05 difference in the LPIPS of the median frame reconstruction and a nearly 0.1 difference between \TheSystem and Pure Upsampling at the tail in \Fig{model_ablation_cdf}. Since one of our goals is to have a robust neural compression system that has good reconstruction quality across a wide variety of frames, this difference at the tail across approaches is particularly salient. On the other hand, \TheSystem synthesizes the mic better, and has fewer blocky artifacts on the face. 

\begin{table}[!t]
    \centering
    \resizebox{\linewidth}{!}{
    \begin{tabular}{lccccc}
    \toprule
    Model Architecture & PSNR (dB) & SSIM (dB) & LPIPS \\
        \midrule
        Upsampling & 24.72 & 7.02  & 0.30 \\
\TheSystem & \textbf{24.90} & \textbf{7.42} &  \textbf{0.26} \\
    \bottomrule
    \end{tabular}
    }
    \caption{Average visual quality of synthesized frames from \TheSystem and ``Pure Upsampling'' when reconstructing from decompressed 128$\times$128 frames.}
    \label{tab:model_architecture_comparison}
    \vspace{-5pt}
\end{table}

\NewPara{Personalization.} To quantify the impact of personalization, we compare two versions of \TheSystem and Pure Upsampling: a \emph{generic model} trained on a large corpus of 512$\times$512 videos~\cite{maxine} of different people and a personalized model fine-tuned on videos of only a specific person. We compare the visual metrics averaged across all videos of all people when synthesized from a single generic model against the case when each person's videos are synthesized from their specific model. All models upsample 64$\times$64 LR frames (without codec compression) to 512$\times$512. \Fig{personalization vs generic 512} visually shows that the personalized \TheSystem model reconstructs the details of the face (row 1), dimples (row 2) and rim of the glasses (row 3) better but \Tab{personalization} shows that personalization provides limited benefits for \TheSystem (PSNR and SSIM improvements of \SI{1.1}{dB} and \SI{0.56}{dB}). However, personalization provides more pronounced benefits of nearly \SI{1.5}{dB} in SSIM and PSNR, and 0.04 in LPIPS for the Pure Upsampling architecture. This suggests that the benefits from personalization are most notable when the underlying model architecture itself is limited in its representational power: the personalized pure upsampling architecture is able to encode high-frequency information in its weights when trained in a personalized manner, while it has no pathways to obtain that information in a generic model. It also illustrates that \TheSystem is more robust; it operates better than pure upsampling in the harder settings of a generic model and extreme upsampling from a 128$\times$128 frame compressed to \SI{15}{Kbps} (\Fig{model_ablation_cdf}).

\begin{table}[!t]
    \centering
    \small
    \resizebox{\columnwidth}{!}{
    \begin{tabular}{llccc}
    \toprule
   Model & Training Regime & PSNR (dB) & SSIM (dB) & LPIPS \\
        \midrule
\multirow{2}*{\textbf{\TheSystem}} & Personalized & \textbf{30.59} & \textbf{11.11} & \textbf{0.13} \\
& Generic & 29.48 & 10.55 & 0.13  \\
        \midrule
\multirow{2}*{\textbf{Upsampling}} & Personalized & \textbf{30.32} & \textbf{11.10} & \textbf{0.13} \\
& Generic & 28.89 & 9.59 & 0.17  \\
    \bottomrule
    \end{tabular}}
    \caption{\small Impact of personalization on different models.}
    \label{tab:personalization}
\end{table}

\subsection{Operational Considerations}
\label{sec:eval_config}
\begin{table}[t] 
    \centering
    \small
    \resizebox{\linewidth}{!}{
    \begin{tabular}{lcccccc}
    \toprule
    \textbf{Convolution Type} & \multicolumn{3}{c}{\textbf{Regular}} & \multicolumn{3}{c}{\textbf{Depthwise}} \\
    \cmidrule(rl){1-1} \cmidrule(rl){2-4} \cmidrule(rl){5-7}
    \textbf{NetAdapt Target} & None & 10\% & 1.5\%
     & None & 10\% & 1.5\% \\
        \midrule
        \# of Decoder MACs & 195B & 14.7B & 2.4B & 22.7B & 1.7B & 0.3B\\
\# of Decoder Parameters & 30M & 1.3M & 151K & 3.4M & 186K & 26K\\
A100 Inference & 13ms & 9ms  & 8ms & 11ms & 8ms & 7ms \\
V100 Inference & 14ms & 11ms & 8ms & 14ms & 8ms & 6ms \\
Titan X Inference & 46ms & 17ms & 12ms & 53ms & 26ms & 13ms  \\
Jetson TX2 Inference & 800ms & 274ms & 165ms & 436ms & 183ms & 87ms\\
LPIPS (Generic) & 0.13 & 0.15  & 0.17 & 0.14 & 0.16 & 0.19 \\
LPIPS (Personalized) & 0.13 & 0.13  & 0.15 & 0.14 & 0.14 & 0.17 \\
    \bottomrule
    \end{tabular}}
    \caption{Accuracy and compute overheads of different versions of our model operating on 512$\times$512 resolution.}
    \label{tab:compute_optimization}
    \vspace{-5pt}
\end{table}



\NewPara{Computational Overheads.}
\Tab{compute_optimization} explores the accuracy \vs compute tradeoffs of different models at 512$\times$512 resolution with and without personalization as measured on different GPU systems with the various optimizations described in \Sec{sec:system optimization compute}. We focus on optimizing the decoding layers (bottleneck and up-sampling) of the generator (~\Fig{neural encoder decoder}) since they are the most compute intensive layers, and are also run on every received frame. The accuracy is reported in the form of the average LPIPS~\cite{lpips} across all frames of our corpus. We observe that DSC reduces the decoder to 11\% of its original MACs. While this gives limited improvements on large GPU systems, it improves the inference time on Jetson TX2, an embedded AI device, by 1.84$\times$. Running NetAdapt further reduces the inference time to \SI{87}{ms} at 1.5\% of the model MACs on the TX2. 
The NVIDIA compiler on the Titan X GPU and the Jetson TX2 is not optimized for DSC~\cite{Wofk2019FastDepthFM}; this can be improved with a TVM compiler stack~\cite{tvm} and optimized engines such as TensorRT~\cite{tensorrt}. However, running NetAdapt produces a real-time model for Titan X even at 10\% of the original model MACs. As expected though, there is a loss of accuracy as the models become smaller. This loss is negligible in moving from the full model MACs to 10\%, particularly when personalizing, but is more significant at 1.5\%. The trend with personalization is expected since smaller models do not generalize well with their limited capacity, however it does not help if the optimizations are extreme. This illustrates that there is a sweet spot (such as decreasing MACs to 10\%) wherein the gains from decreased compute outweigh the loss (or lack thereof) in accuracy. 

\NewPara{Choosing \PF Stream Resolution.}
\TheSystem is designed flexibly to work with LR frames of any size (64$\times$64, 128$\times$128, 256$\times$256, 512$\times$512) to resolve them to 1024$\times$1024 frames, and to fall back to VPX at full resolution if it can be supported. VP8 and VP9 achieve different bitrate ranges at every resolution by varying how the video is quantized. For instance, on our corpus, we observe that 256$\times$256 frames can be compressed with VP8 in the \SI{45}{Kbps}--\SI{180}{Kbps} range, but VP9 can compress even 512$\times$512 frames from \SI{75}{Kbps} onwards. These bitrate ranges often overlap partially across resolutions. This begs the question: given a target bitrate, what resolution and codec should the model use to achieve the best quality? To answer this, we compare the synthesis quality with \TheSystem atop VP8 from three \PF resolutions, all at \SI{45}{Kbps} in \Tab{resolution_comparison}. Upsampling 256$\times$256 frames, even though they have been compressed
more to achieve the same bitrate, gives a nearly \SI{4}{dB} improvement in PSNR, more than \SI{2}{dB} improvement in SSIM, and a 0.03 improvement in LPIPS, over upsampling lower resolution frames. This is because the extent of super-resolution that the model performs decreases dramatically at higher starting resolutions. This suggests that for any given bitrate budget, 
we should start with the highest resolution frames that the \PF stream supports at that bitrate, even at the cost of more quantization. This also means that if VP9 can compress higher resolution frames than VP8 at the same target bitrate, we should pick VP9. \Tab{resolution_bitrate_regime} shows the resolution and codec we choose for different target bitrate ranges in our implementation.

\begin{table}[!t]
    \centering
    \small
    \resizebox{\columnwidth}{!}{
    \begin{tabular}{lccccc}
    \toprule
    \PF Stream Resolution & PSNR (dB) & SSIM (dB) & LPIPS \\
        \midrule
        64$\times$64 & 23.80 & 6.77 & 0.27 \\
        128$\times$128 & 25.72 & 7.86 & 0.27   \\
 256$\times$256 & \textbf{27.12} & \textbf{9.01} & \textbf{0.24}  \\
    \bottomrule
    \end{tabular}}
    \caption{Reconstruction quality from different resolution \PF stream frames at the same bitrate of 45 Kbps. \TheSystem reconstructs better from higher resolution frames.}
    \label{tab:resolution_comparison}
    \vspace{-15pt}
\end{table}

\NewPara{Encoding Video During Training.} A key insight in the design of \TheSystem is that we need to design the neural compression pipeline to leverage the latest developments in codec design. One way to do so is to allow the model to see decompressed frames at the chosen bitrate and \PF resolution during the training process so that it learns the artifacts produced by the codec. This allows us to get extremely low bitrates for LR frames (which often causes color shifts or other artifacts) while maintaining good visual quality. To evaluate the benefit of this approach, we compare five training regimes for \TheSystem when upsampling 128$\times$128 video to 1024$\times$1024: (1) no codec, (2) VP8 frames at \SI{15}{Kbps}, (3) VP8 frames at \SI{45}{Kbps}, (4) VP8 frames at \SI{75}{Kbps}, (5) VP8 frames at a bitrate uniformly sampled from \SI{15}{Kbps} to \SI{75}{Kbps}. We evaluate all five models at upsampling decompressed frames at \SI{15}{Kbps}, \SI{45}{Kbps}, and \SI{75}{Kbps}. 

\begin{table}
    \centering
    \resizebox{\columnwidth}{!}{
    \begin{tabular}{lccccc}
    \toprule
    Training Regime & \PF @ 15 Kbps & \PF @45 Kbps & \PF @75 Kbps \\
        \midrule
        No Codec & 0.32 & 0.30 & 0.28 \\
        VP8 @ 15 Kbps & \textbf{0.26} & \textbf{0.25} & \textbf{0.23}  \\
        VP8 @ 45 Kbps & 0.28 & 0.27 & 0.25  \\
        VP8 @ 75 Kbps & 0.30 & 0.28 & 0.26  \\
        VP8 @ [15, 75] Kbps & 0.28 & 0.26 & 0.25   \\
    \bottomrule
    \end{tabular}}
    \caption{LPIPS for different regimes wherein we include the VP8 codec in the training pipeline. The model trained with the lowest bitrate videos at a given resolution performs best regardless of what the bitrate of the video is at inference time.}
    \label{tab:encoder_effect}
    \vspace{-14pt}
\end{table}

\Tab{encoder_effect} shows the LPIPS achieved by all the models in each reconstruction regime. All models trained on decompressed frames perform better than the model trained without the codec. 
Further, the model trained at the lowest bitrate (\SI{15}{Kbps}) performs the best even when provided decompressed frames at a higher bitrate at test time because it has learned the most challenging Super-Resolution task from the worst LR frames, and performs
well even with easier instances or higher bitrate frames. 
This suggests that we only need to train one personalized model per \PF resolution at the lowest bitrate supported by a resolution, and then we can reuse it across the entire bitrate range that the \PF resolution can support.



\subsection{Adaptation to Network Conditions}
\label{sec:eval_adaptation}

\begin{figure}[!t]
\centering
\includegraphics[width=0.75\linewidth]{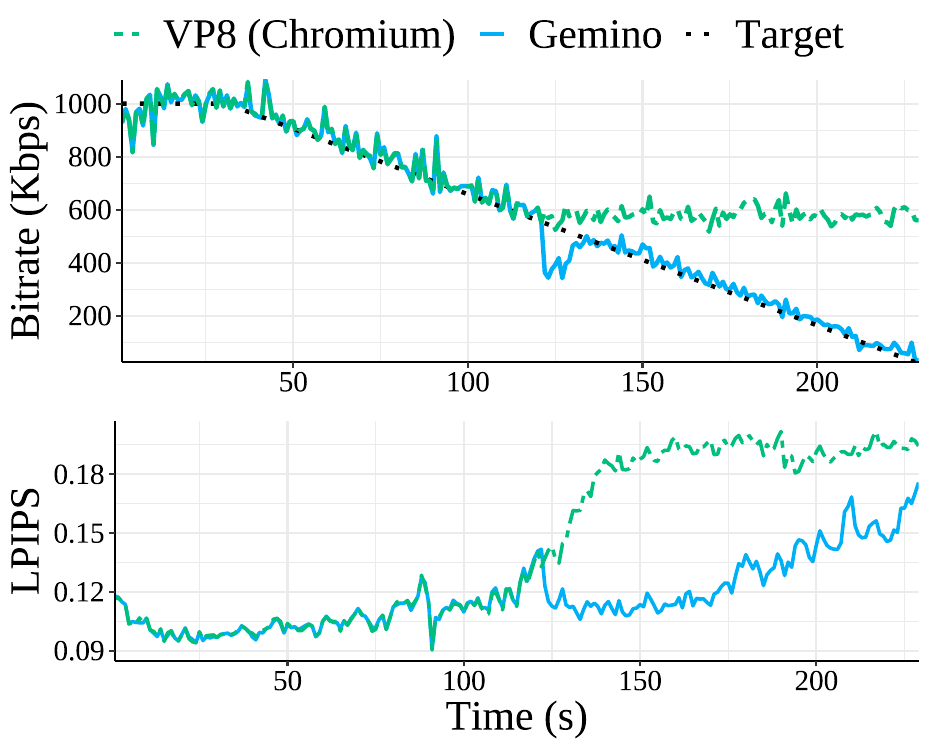}
\caption{\small \TheSystem's ability to adapt to a time-varying target bitrate. As the target bitrate reduces, \TheSystem gradually lowers its \PF stream resolution trading off more upsampling and less quality (increased LPIPS) for a reduction in achieved bitrate. VP8, in contrast, lowers the bitrate initially, but once at its minimum quality, it stops responding to the target bitrate.}
\label{fig:adaptation}
\vspace{-5pt}
\end{figure}

To understand the adaptability of \TheSystem, we explore how it responds to changes in the target bitrate over the course of a video. We remove any conflating effects from bandwidth prediction by directly supplying the target bitrate as a decreasing function of time to both \TheSystem and the VP8 codec. \TheSystem uses only VP8 through all bitrates for a fair comparison.
\Fig{adaptation} shows (in black) the target bitrate, along with the achieved bitrates of both schemes, and the associated perceptual quality~\cite{lpips} for a single video over
the course of 220s of video-time\footnote{The timeseries are aligned to ensure that VP8 and \TheSystem receive the same video frames to remove confounding effects of differing latencies.}. We observe that initially (first 120s), at high target bitrates, \TheSystem and VP8 perform very similarly because they are both transmitting
just VP8 compressed frames at full (1024$\times$1024) resolution. 
Once VP8 has hit its minimum achievable bitrate of $\sim$550~Kbps (after 120s), there is nothing more it can do, and it stops responding to the input target bitrate. However, \TheSystem continues to lower its \PF stream resolution and/or bitrate in small steps all the way to the lowest target bitrate of \SI{20}{Kbps}.  Since \TheSystem is only using VP8 here, it switches to 512$\times$512 at \SI{550}{Kbps}, 256$\times$256 at \SI{180}{Kbps}, and 128$\times$128 at \SI{30}{Kbps}.
This design choice might cause abrupt shifts in quality around the transition points between resolutions. However, \TheSystem prioritizes responsiveness to the target bitrate over the hysteresis that classical encoders experience which, in turn, leads to packet losses due to overshooting and glitches.  As the resolution of the \PF stream decreases, as expected, the perceptual quality of \TheSystem worsens but is still better than VP8's visual quality. This shows that \TheSystem can adapt well to bandwidth variations, though we leave the design of a transport and adaptation layer that provides fast and accurate feedback to \TheSystem for future work. 


\section{Limitations and Future Work}
\label{sec:limitations}
While \TheSystem greatly expands the operating regime for video conferencing to very low bitrates, it incurs significant overheads in the form of training costs for codec-in-the-loop training and personalization. It compresses better than VPX, but the encoding and decoding processes are quite a bit slower than VPX, and not as widely supported on devices without access to some graphical processing engine. However, we believe that device improvements year on year are trending in a favorable direction, particularly with the emergence of optimized runtimes and hardware for running machine learning workloads on both Apple and Android devices~\cite{apple_neural_engine, android_nnapi}. Further, NetAdapt~\cite{eccv2018-yang} and layer-by-layer pruning is only one technique amongst a large suite of model optimization approaches. We believe that with more targeted optimizations for particular devices, we can do better. Such optimizations become more salient when operating on higher-resolution video (\eg 4K, UltraHD) and in higher bandwidth regimes ($\sim$ \SI{5}{Mbps}). We leave an exploration of such optimizations to future work. 

\TheSystem, though trained on random pairs of reference and target frames, always uses the first frame of the test video as its reference frame. The reconstruction fidelity can be improved by using reference frames close to each target frame. However, sending more frequent reference frames incurs very high bitrate costs due to their high resolution. We leave to future work a more thorough investigation of reference frame selection mechanisms that weigh these tradeoffs to squeeze the maximum accuracy for a given compression level.

%

\section{Conclusion}
\label{sec:conclusion}
This paper proposes \TheSystem, a neural video compression scheme for \vcing using a new high-frequency-conditional super-resolution model. Our model combines the benefits of low-frequency reconstruction from a low-resolution target, and high-frequency reconstruction from a high-resolution reference. Our novel \ms architecture and personalized training synthesize good quality videos at high resolution across many scenarios. The adaptability of the compression scheme to different points on a rate-distortion curve opens up new avenues to co-design the application and transport layers for better quality video calls. However, while neural compression shows promise in enabling very low bitrate video calls, it also raises important ethical considerations about the bias that training data can introduce on the usefulness of such a technique to different segments of the human population. We believe that our personalized approach alleviates some of these concerns, but does not eliminate them entirely.%
\label{beforerefs} 



\section*{Acknowledgments}
  We thank our shepherd, Arpit Gupta, and our anonymous NSDI reviewers for their feedback. This work was supported by
  GIST, seed grants from the MIT Nano NCSOFT program and Zoom Video Communications, and NSF awards 2105819, 1751009 and 1910676. 

\bibliographystyle{unsrt}
\bibliography{reference}

\newpage
\appendix
\noindent

\section{Model Details}
\label{app:model details}
In the following subsections, we detail the structure of the motion estimator that produces the warping field for \TheSystem and the neural encoder-decoder pair that produce the prediction. We also describe additional details about the training procedure.

\subsection{Motion Estimator}
\label{sec:app:motion_estimator}

\begin{figure}
\centering
\includegraphics[width=\linewidth]{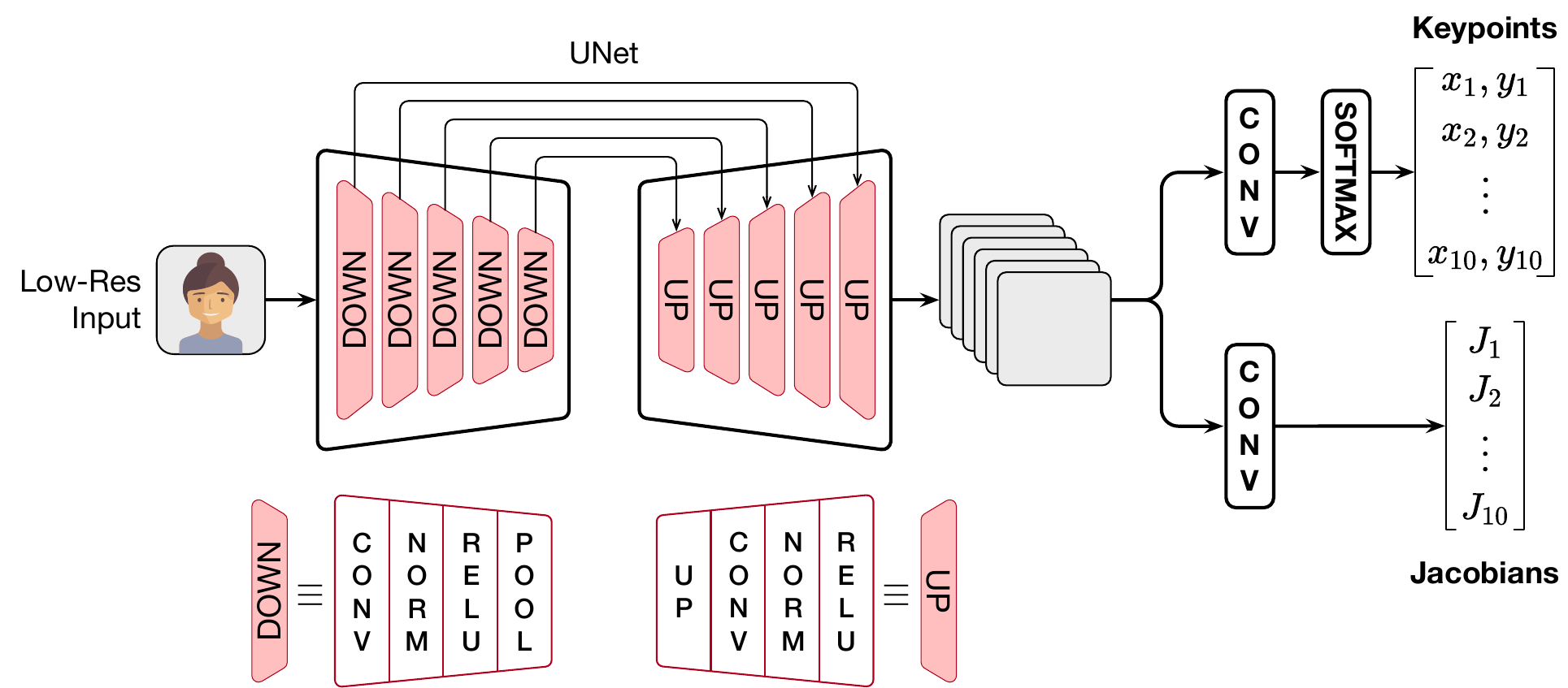}
\caption{\small Keypoint Detector used as a precursor for computing the warping field between the reference and target images. Low-resolution versions of both frames are supplied to a UNet architecture~\cite{unet}, and then put through convolutional layers to generate \kp locations and four ``Jacobian'' values in the neighborhood of each \kp.}
\label{fig:kp detector}
\end{figure}

\begin{figure*}[h]
\centering
\includegraphics[width=\linewidth]{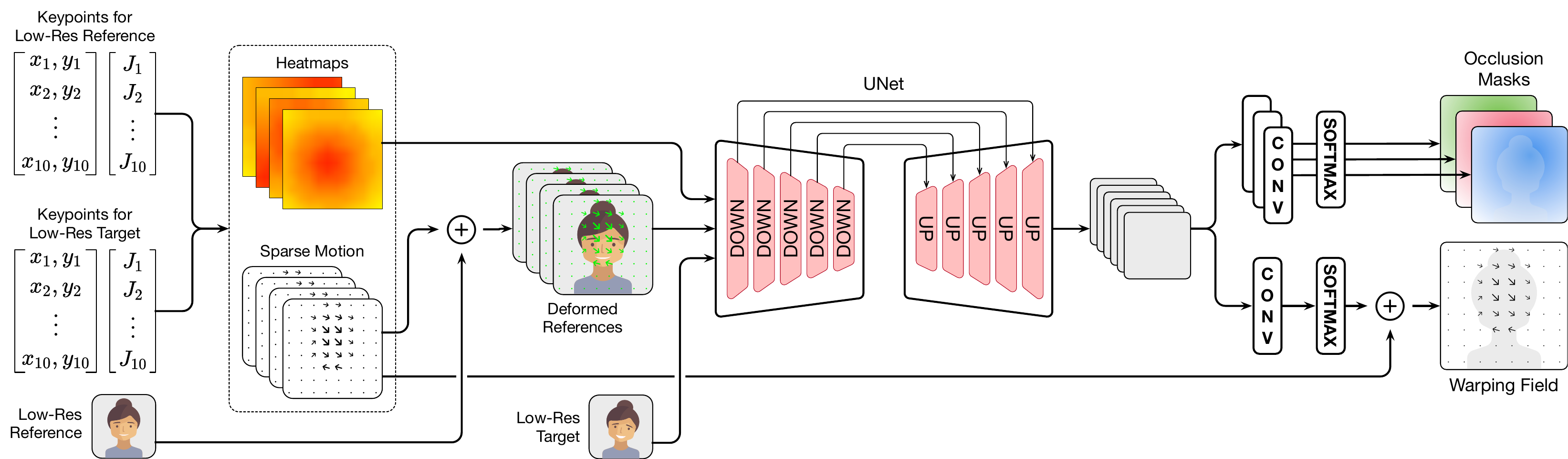}
\caption{\small \TheSystem's motion estimation module that takes as input reference and target \kps, along with a low-resolution reference and target frames, and produces a warping field and  occlusion masks. The warping field helps the generator move encoded features from a full-resolution reference frame into the target frame's coordinate space, while the occlusion masks inform the decoder how to combine information from the low-resolution target frame with high-resolution (warped and unwarped) features from the reference frame.}
\label{fig:motion estimator}
\end{figure*}

\NewPara{UNet Structure.}
The \kp detector and the motion estimator use identical UNet structures (\Fig{kp detector} and \Fig{motion estimator}) to extract features from their respective inputs before they are post-processed. In both cases, the UNet consists of five up and downsampling blocks each. Each downsampling block consists of a 2D convolutional layer~\cite{conv}, a batch normalization layer~\cite{batchnorm}, a Rectified Linear Unit Non-linearity (ReLU) layer~\cite{relu}, and a pooling layer that downsamples by 2$\times$ in each dimension. The batch normalization helps normalize inputs and outputs across layers, while the ReLU layer helps speed up training. Each upsampling block first performs a 2$\times$ interpolation, followed by a convolutional layer, a batch normalization layer, and a ReLU layer. Thus, every downsampling layer reduces the spatial dimensions of the input but instead extracts features in a third ``channel'' or ``depth'' dimension by doubling the third dimension. On the other hand, every upsampling layer doubles in each spatial dimension, while halving the number of
features in the depth dimension. In our implementation, the UNet structure always produces 64 features after its first encoder downsampling layer, and doubles from there on. The reverse happens with the decoder ending with 64 features after its last layer. 
Since the UNet structure operates on low-resolution input (as part of the \kp detector and motion estimator), its kernel size is set to 3$\times$3 to capture reasonably sized fields of interest.

\NewPara{Keypoint Detector.}
To obtain the warping field between the reference frame and the target, \TheSystem first uses a \kp detector to locate key facial landmarks. It then uses a first-order approximation in the neighborhood of these \kps similar to the FOMM~\cite{fom}. To extract \kps, we first downsample the input image to 64$\times$64, and then feed it into the UNet structure described above in its RGB space itself. The UNet structure produces
a set of output features from its decoder, which are then put
through two separate pipelines to extract the \kp locations and the ``Jacobians.'' The \kp locations are extracted via a single 7$\times$7 convolutional layer, which is then put through a softmax to extract probabilities for \kp presence at each spatial location. 
This is then converted to actual \kp locations by performing a weighted average of these probabilities across the entire spatial grid. Note that this process is replicated 10 times by having 10 separate channels to extract 10 \kps. The Jacobians are simply four floating point numbers that are used to approximate the movement (derivatives) in the neighborhood of each \kp. This is used for
the first-order approximation when computing the motion around each \kp. To generate these Jacobians, the output from the UNet is simply put through a single 7x7 convolutional layer. \Fig{kp detector} describes this architecture. Note that both the reference and the target images are fed to this pipeline independently to generate
two separate sets of reference and target \kps and Jacobians.

\NewPara{Motion Estimation}
\Fig{motion estimator} describes the working of the motion estimator in \TheSystem's design in detail. First, the motion estimator creates Gaussian \emph{heatmaps} corresponding to the \kp locations from both the reference and the target frames. It subtracts the two on a per-\kp basis to generate the difference between the two frames' \kps. It adds a separate heatmap consisting of zeros to denote the fact that the background is identical in the two frames. The motion estimator then generates \emph{sparse motion} vectors or motion vectors in the neighborhood of each \kp using the first-order Taylor series approximation~\cite{fom} and the Jacobian values from the \kp detector. These motion vectors (along with an identity for the background) are applied to the low-resolution reference frame to obtain a set of \emph{deformed references}. This effectively generates 11 heatmaps (10 \kps + 1 for background), and 11 different RGB (3 channels) deformed references. The 44 resulting channels are provided as input along with 3 RGB features from the low-resolution target image to another replica of the UNet structure described above. This
UNet's decoder also outputs a set of predicted features based on all the provided 47 input features. 

The predicted features are put through three separate 7$\times$7 convolutional layer followed by Sigmoid layers and a Softmax layer to produce \emph{three occlusion masks}.
Each occlusion mask is later used in the decoding pipeline to convey how to combine information from three pathways: the warped high-resolution features, the non-warped high-resolution features, and the low-resolution features to generate the prediction. We use a Softmax layer to enforce that the sum of these three occlusion masks is 1 at every spatial location 
so that they do not compete in later parts of the decoding pipeline. Intuitively, this forces each pixel to be generated from one out of the three pathways. If a feature represents a part of the frame that has moved between the reference and the target frames, reconstruction relies on the HR warped pathway, while if it represents a part of the frame that has not moved, it relies on the non-warped HR pathway. Regions that are significantly different between the reference and the target use the LR features instead.

The predicted features are also fed into a single 7$\times$7 convolutional layer followed by a Softmax to generate a \emph{deformation mask} that reflects how to combine the sparse motion vectors previously obtained in the neighborhood of each location. The Softmax ensures that across every spatial location, different vectors are weighted appropriately to sum up to 1 finally. This deformation mask is applied to the sparse motion vectors to obtain the final warping field.



\begin{figure*}[h]
\centering
\includegraphics[width=\linewidth]{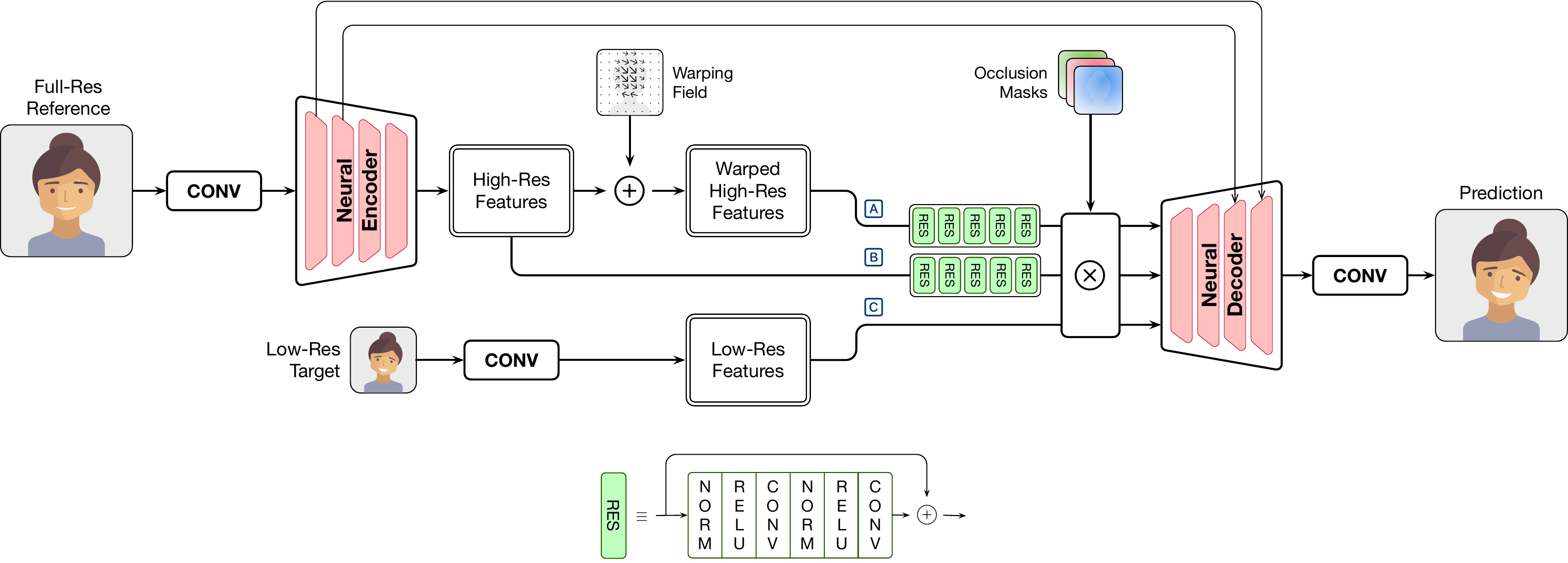}
\caption{\small \TheSystem's encoder-decoder pair that is responsible for synthesizing the prediction. The encoder runs the high-resolution reference image through a series of downsampling layers to produce encoded features. A copy of these encoded HR features is warped, and both the warped and non-warped features are refined through a sequence of residual blocks. Meanwhile, a convolutional layer extracts low-resolution features from a low-resolution target. The three sets of features are combined based on occlusion masks from the motion estimator before they are decoded to result in the final prediction.}
\label{fig:neural encoder decoder}
\end{figure*}

\subsection{Image Synthesis}
\label{sec:app:encoder_decoder}
\Fig{neural encoder decoder} describes the generative parts of \TheSystem design in more detail. First, the low-resolution target frames
are put through a single 7$\times7$ convolutional layer to produce low-resolution (LR) features to be used later during decoding. 
The high-resolution reference RGB frame is also fed through a single 7$\times7$ convolutional layer to produce 32 high-resolution (HR) features with the same spatial dimensions before it is fed to the neural encoder.
The neural encoder consists of four downsample blocks, each of which has the same structure as the downsample blocks in our UNet structure in the \kp detector (\Fig{kp detector}).
The four blocks ensure that we start at full-resolution (1024$\times$1024 frames) with 32 features and end up with 256 separate 64$\times$64 encoded HR features at the bottleneck.
However, unlike the UNet, not all blocks in the neural encoder are equipped with skip connections. 
Specifically, only the first two have skip connections to the corresponding last two decoder blocks. The decoder also consists of upsampling blocks with the
same composition as the UNet’s downsampling blocks. 
Prior to decoding, a copy of the HR features are warped using the warping field from the motion estimator (\sfboxed{A}). The warped HR features and the unwarped features (\sfboxed{B}) are fed through five Residual Blocks~\cite{he2016deep} that refine them and also help prevent diminishing gradients during training. Each residual block consists of batch normalization, ReLU, and convolutional layers. The refined HR features, along with the LR features, and appropriate skip connections are combined using occlusion masks obtained from the motion estimator. Once combined, all three sets of features are fed through decoder's upsampling blocks to spatial dimensions of 1024$\times$1024 frames with 32 features. This output is fed through a final 7$\times$7 convolutional layer to bring it back to its RGB space as the prediction.

\subsection{Training Details}
\label{app:training}
To train our model, we use equally weighted (with a weight 10) multi-scale VGG perceptual loss~\cite{johnson2016perceptual}, a feature-matching loss~\cite{wang2018video}, and an L1 pixelwise loss. The multi-scale VGG loss and feature matching losses operate at scales of 1, 0.5, 0.25, and 0.125 of the image size. All of these scales have equal loss weights. We also use an adversarial loss~\cite{goodfellow2014generative} with a weight of one (one-tenth of the other losses). The \kps use an equivariance loss similar to the FOMM~\cite{fom}. Our \kp detector is unchanged relative to the FOMM, and so we reuse a trained checkpoint on the VoxCeleb dataset~\cite{voxceleb}, but fine-tune it on a per-person basis. For other layers, we adopt a simple strategy: if the dimensions match the equivalent layers in the FOMM (\eg encoder layers, residual layers), we initialize them based on a FOMM checkpoint, and if the dimensions do not match, they are initialized randomly. All models are personalized (either trained from scratch or fine-tuned if the layers are initialized to FOMM checkpoints). All models for 1024$\times$1024 output resolution were trained with a batch size of 2 on NVIDIA A100 GPUs while models for 512$\times$512 output resolution were trained with a batch size of 2 on NVIDIA V100 GPUs. The details of the dataset we curate is shown in \Tab{dataset_info}
As discussed earlier, models are trained on decompressed VPX frames at different bitrates. To support this, we encode individual frames
at the target bitrate during train-time \emph{after} they are sampled from the dataset. Since we encode individual frames, this only results in keyframes at train-time, but these frames are still representative of what compressed video frames more generally look like at that bitrate.

\begin{table}[t] 
    \centering
    \small
    \resizebox{\linewidth}{!}{
    \begin{tabular}{lcccc}
    \toprule
    & \multicolumn{2}{c}{\textbf{Training Videos}} & \multicolumn{2}{c}{\textbf{Test Videos}} \\
    \cmidrule(rl){2-3} \cmidrule(rl){4-5}
    \textbf{Youtuber} & Total Len. & Avg. Bitrate & Avg. Len. & Avg. Bitrate \\
        \midrule
        \href{https://www.youtube.com/c/adamneely}{Adam Neely} & 31 min & 1082 Kbps & 146 s & 1303 Kbps \\
\href{https://www.youtube.com/c/XiranJayZhao}{Xiran Jay Zhao} & 30 min & 2815 Kbps & 180 s & 1560 Kbps \\
\href{https://www.youtube.com/user/theneedledrop}{The Needle Drop} & 33 min & 2013 Kbps & 206 s & 1286 Kbps \\
\href{https://www.youtube.com/c/fancyfueko}{fancy fueko} & 15 min & 4064 Kbps & 124 s & 1607 Kbps \\
\href{https://www.youtube.com/c/PBSNewsHour}{Kayleigh McEnany} & 28 min & 2521 Kbps & 180 s & 1247 Kbps \\
    \bottomrule
    \end{tabular}}
    \caption{Details of our dataset. All videos are at 1024$\times$1024.}
    \label{tab:dataset_info}
    \vspace{-10pt}
\end{table}

\section{Extended Evaluation}
\label{app:extra eval}
\subsection{Low-bitrate Regime Comparisons}
\label{sec:app:low bitrate}
\begin{figure*}[t]
\centering
\subfigure[\small PSNR achieved by different schemes in the low-bitrate regime]{
\centering
\includegraphics[width=0.7\linewidth]{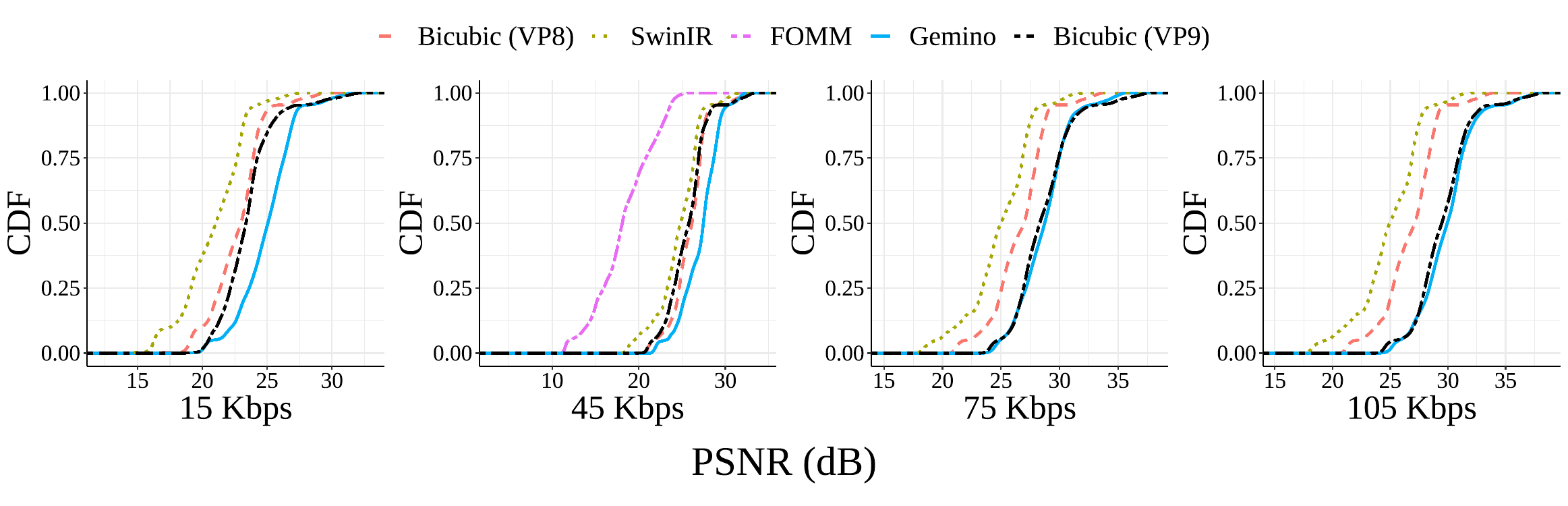}
\label{fig:main_cdf_psnr}
}
\subfigure[\small SSIM achieved by different schemes in the low-bitrate regime]{
\centering
\includegraphics[width=0.7\linewidth]{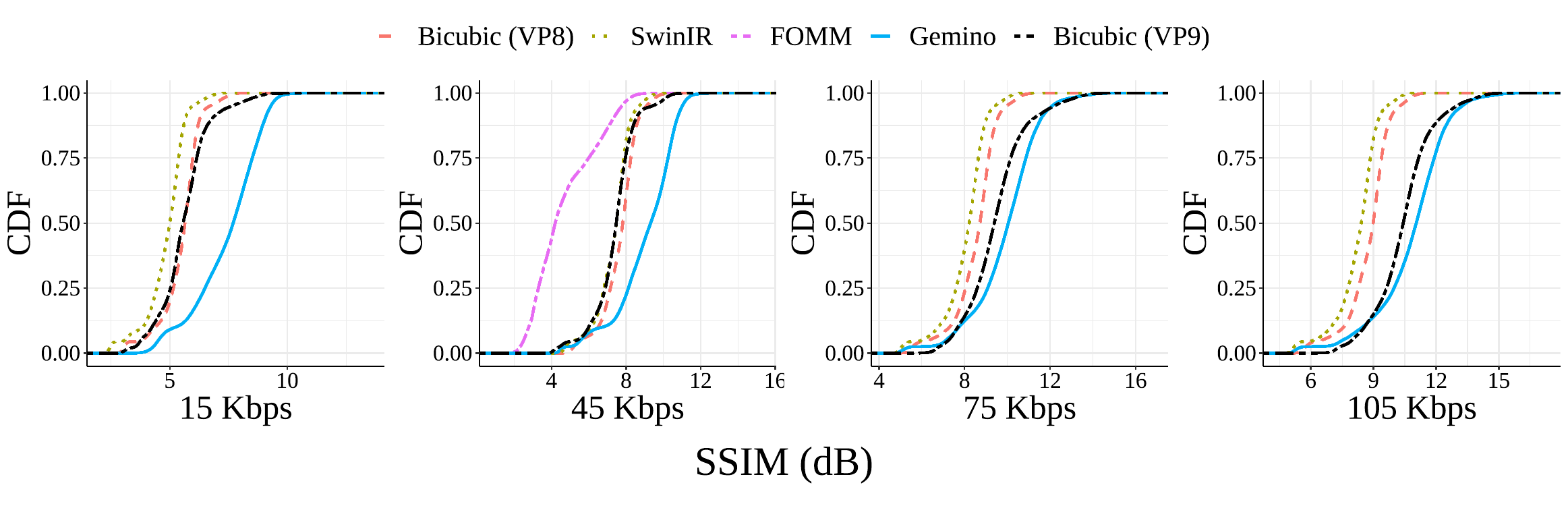}
\label{fig:main_cdf_ssim}
}
\caption{\small CDF of PSNR and SSIM achieved by different schemes on all frames in low-bitrate regimes. The differences between Gemino and other approaches becomes more pronounced in lower-bitrate regimes.}
\label{fig:main_cdf_psnr_ssim}
\vspace{-12pt}
\end{figure*}

We plot a CDF of the visual quality in the context of PSNR and SSIM across all frames in our corpus in \Fig{main_cdf_psnr_ssim} at \SI{15}{Kbps}, \SI{45}{Kbps}, \SI{75}{Kbps} and \SI{105}{Kbps}. At each target bitrate, we use the largest resolution supported by the underlying video codec. \TheSystem uses VP8 at \SI{15}{Kbps} and \SI{45}{Kbps} but VP9 at higher bitrates; VP8 and VP9 do not differ much at lower bitrates, but VP9 allows 512$\times$512 to be compressed all the way to \SI{75}{Kbps} while VP8 cannot. Since our results in \Sec{sec:eval_config} suggest using the highest resolution at any target bitrate, we use VP9 at \SI{75}{Kbps} and \SI{105}{Kbps}.
The CDF shows that \TheSystem's reconstructions are robust to variations across frames and orientation changes over the course of a video.
At \SI{45}{Kbps} when upsampling from a 256$\times$256 frame, \TheSystem outperforms all other baselines across all frames and both metrics. Specifically, its synthesized frames are better than FOMM 
by nearly \SI{5}{dB} in SSIM and \SI{10}{dB} in PSNR throughout. While the differences in PSNR between Bicubic atop VP9 and \TheSystem are small at \SI{105}{Kbps}, the difference at the median is over \SI{2}{dB} at \SI{15}{Kbps}. The differences in SSIM is even more pronounced at \SI{15}{Kbps} with \TheSystem outperforming both bicubic approaches by nearly \SI{2.5}{dB} at the median. As the bitrate is lowered, the differences between VP8 and VP9 also shrink, resulting in very similar visual quality for both of their bicubic upsampling approaches. As we show in \Sec{sec:app:metrics}, differences in PSNR are not as reflective of how natural the image looks to the human eye, and so we rely on LPIPS~\cite{lpips} primarily, and SSIM when relevant.
\begin{figure*}
\centering
\includegraphics[width=\linewidth]{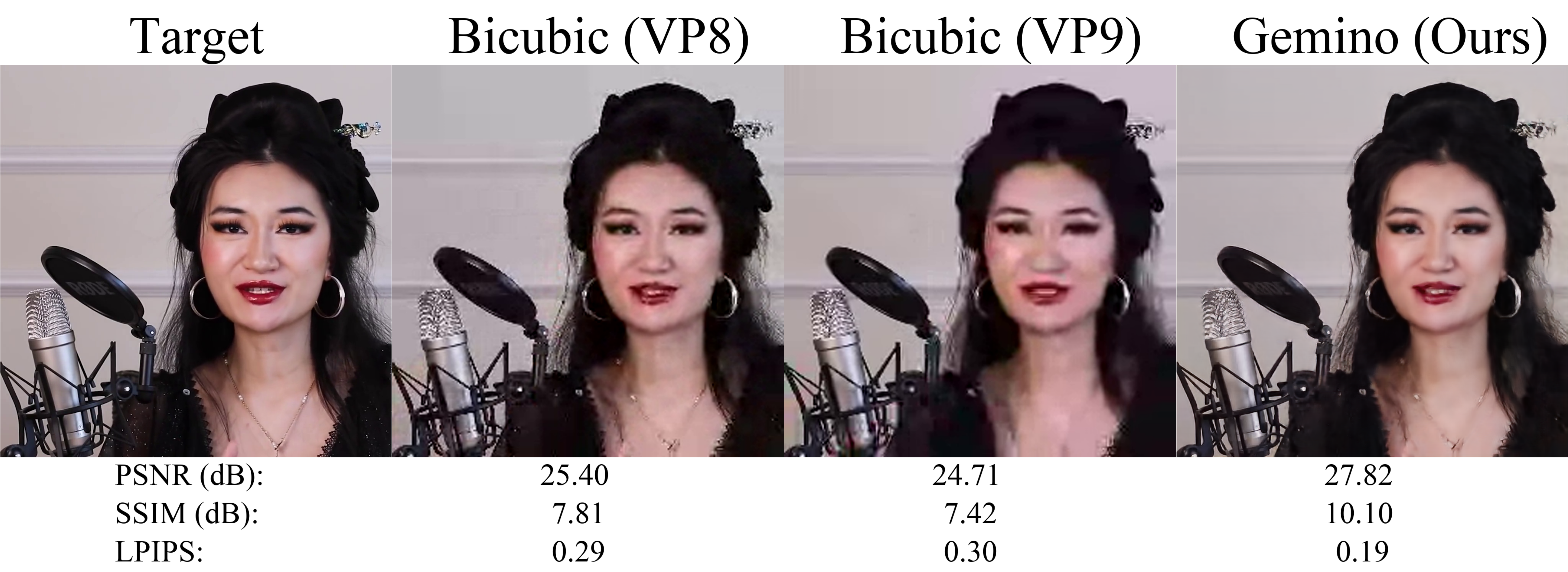}
\caption{\small Full-size examples to illustrate the visual quality differences and their correlation with the metric values.}
    \label{fig:metrics comparison}
\end{figure*}

\subsection{Comparing the metrics.}
\label{sec:app:metrics}
\begin{figure*}[h]
\centering
\includegraphics[width=\linewidth]{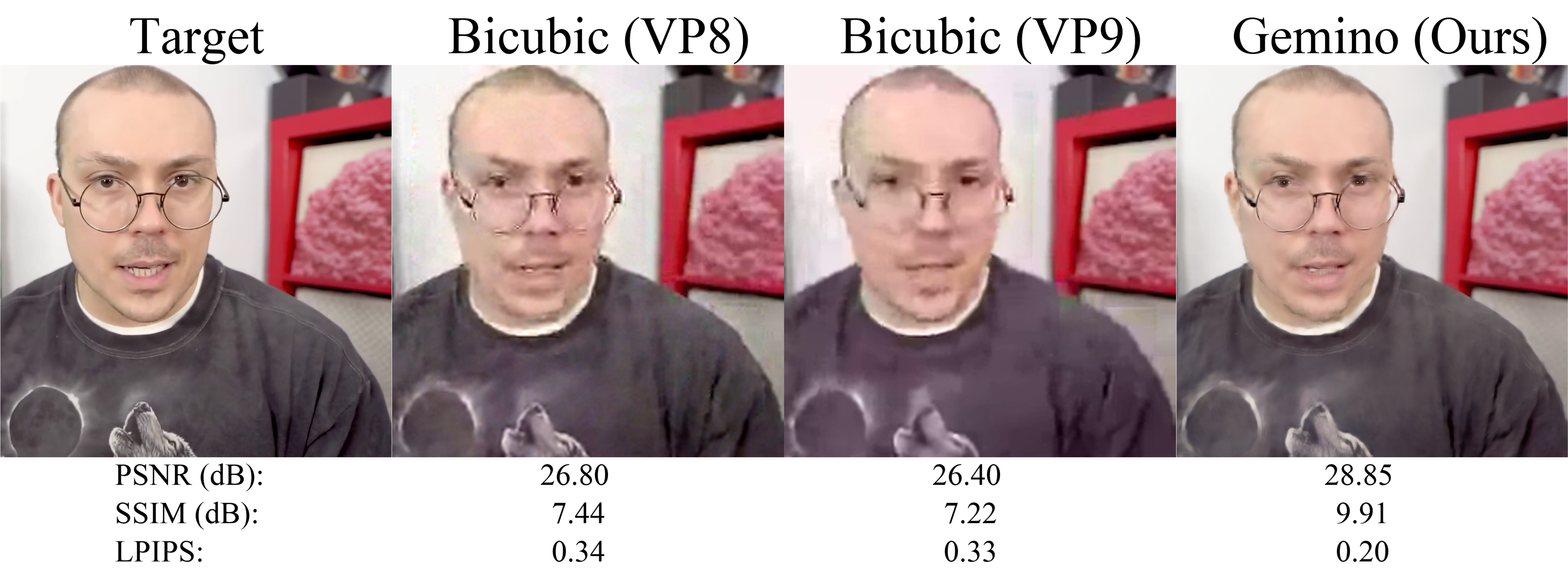}
\caption{\small Full-size examples to illustrate the visual quality differences and their correlation with the metric values.}
    \label{fig:metrics comparison 2}
\end{figure*}

\begin{figure*}[h]
\centering
\includegraphics[width=\linewidth]{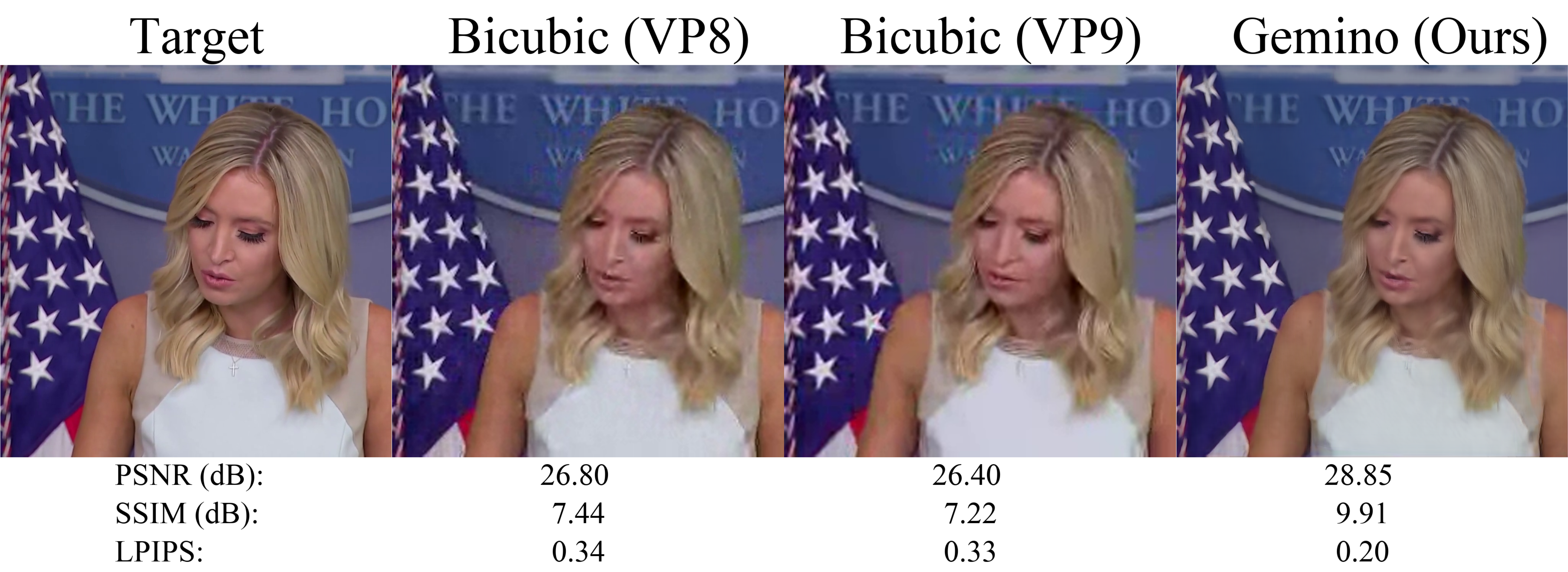}
\caption{\small Full-size examples to illustrate the visual quality differences and their correlation with the metric values.}
    \label{fig:metrics comparison 3}
\end{figure*}

\begin{figure*}[h]
\centering
\includegraphics[width=\linewidth]{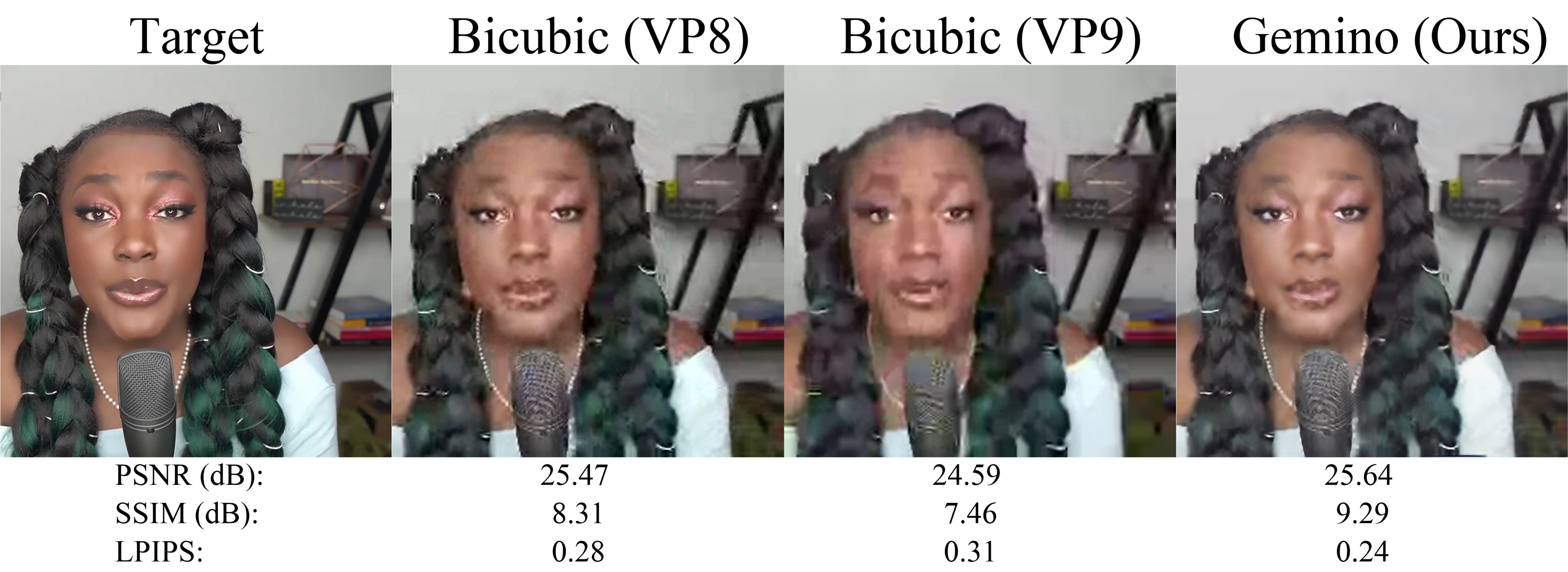}
\caption{\small Full-size examples to illustrate the visual quality differences and their correlation with the metric values.}
    \label{fig:metrics comparison 4}
\end{figure*}

To understand the difference between different quantitative metrics and the visual quality each of them map to, we show a strip (with the associated metrics) that compares \TheSystem with the reconstruction of bicubic atop VP8 and VP9 when upsampling 256$\times$256 frames at \SI{45}{Kbps} in \Fig{metrics comparison}, ~\ref{fig:metrics comparison 2}, ~\ref{fig:metrics comparison 3} and ~\ref{fig:metrics comparison 4}. For example, in \Fig{metrics comparison}, while the output produced by \TheSystem is significantly better than that of both bicubic approaches, it manifests as only a \SI{2}{dB} difference in PSNR and SSIM, both of which have very wide ranges. In contrast, we see an improvement of $0.1$ in LPIPS. LPIPS is constrained to be between 0 and 1, making it easier to put the 0.1 improvement in context. We see similar trends in \Fig{metrics comparison 2} and in \Fig{metrics comparison 3}, where a difference of 0.13-0.14 in LPIPS, and $\sim$\SI{2}{dB} PSNR manifests as smoother output on the facial regions. In contrast, while the PSNRs in \Fig{metrics comparison 4} are fairly close across the bicubic approaches and \TheSystem, the SSIM and LPIPS differences reflect how much more natural \TheSystem's facial reconstruction looks. Frames like these explain why the PSNR graphs in \Fig{main_comparison} are a lot closer than the other two metrics, and motivate the use of LPIPS as our main metric.  However, even though  the two bicubic approaches in \Fig{metrics comparison 2} differ significantly in the smoothness of the face, all three metrics are close enough that the difference would not be very perceptible on a plot. As a result, we acknowledge that none of these metrics are truly perfect, and a combination of them all along with visual strips is needed for a thorough evaluation. 

\end{document}